\documentclass[twocolumn,showpacs,preprintnumbers,amsmath,amssymb]{revtex4}

\usepackage{color}
\usepackage{graphicx}
\usepackage{dcolumn}
\usepackage{bm}
\usepackage{multirow}
\usepackage{multirow}

\definecolor{sienna}{cmyk}{0,0.72,1,0.45}
\definecolor{fg}{cmyk}{0.91,0,0.88,.12}
\definecolor{yellow}{cmyk}{0,0,1,0}
\definecolor{or}{cmyk}{0,1,0.5,0}
\definecolor{magenta}{cmyk}{0,1,0,0}
\definecolor{rubinered}{cmyk}{0,1,0.13,0.45}
\definecolor{blue}{cmyk}{1,1,0,0}
\definecolor{turquoise}{cmyk}{1,1,0,0.5}
\definecolor{aquamarine}{cmyk}{0,1,0,0.0}
\definecolor{midnightblue}{cmyk}{1,0.5,0.0,0.0}
\definecolor{junglegreen}{cmyk}{1,0,0.2,0.5}

\begin{document}

\title{Queue Length Synchronization in a Communication Network}
                                                              
\author{Satyam Mukherjee}                                     
\email{mukherjee@physics.iitm.ac.in}                          
\author{Neelima Gupte}                                        
\email{gupte@physics.iitm.ac.in}                              
\affiliation{Department of Physics, Indian Institute of Technology
Madras, Chennai - 600036, India.}                             
                                                              
\date{\today}                                                 
                                                              
\begin{abstract}                                              
We study synchronization in the context of network traffic on a $2-d$ communication network with local clustering and geographic separations.  The  network consists of nodes and randomly  distributed hubs where the top five hubs ranked
according to their coefficient of betweenness centrality (CBC) are connected by random assortative and gradient mechanisms. For multiple message traffic, messages can trap at the high
CBC hubs, and congestion can build up on the network with long queues at the congested hubs. The queue lengths are seen to synchronize in the congested phase. Both complete and phase synchronization is seen, between pairs of hubs. In the decongested phase, the pairs start clearing, and synchronization is
lost. A cascading master-slave relation is seen between the hubs, with the  slower hubs (which are slow to decongest) driving the faster ones. These are usually the hubs of high CBC.  Similar results are seen for traffic of constant density. Total synchronization between the hubs of high CBC is also seen in the congested regime. Similar behavior is seen for traffic on a network constructed using the Waxman random topology generator. We also demonstrate the existence of phase synchronization in real Internet traffic data.                                                             
\end{abstract}

\maketitle                          

\section{Introduction}

The phenomenon of synchronization has been studied in
contexts ranging from the synchronization of clocks and the      
flashing of fire-flies \cite{synch} to synchronization in oscillator networks \cite{carol} and in complex
networks \cite{kurths}.  Synchronized states have been seen  in the context of traffic flows as well \cite{kerner}, and
investigations of traffic flow on substrates of various geometries have been the focus of recent research interest \cite{tadic,wang,Jiang,moreno}.  The
synchronization of processes at the nodes, or hubs, of complex networks can have serious 
consequences for the performance of the network \cite{pipa}. In the case of communication networks, 
the performance of the networks is assessed 
in terms of their efficiency at packet delivery. 
Such networks can show a congestion-decongestion transition \cite{congest}.
We note that an intimate connection between congestion and
synchronization effects has been seen in the case of real networks \cite{TCP,huang}.

The aim of this paper is to study the interplay of congestion and synchronization effects on each other, and examine their effect on the efficiency of the network for packet delivery in the context of two model networks based on two dimensional grids. The first network consists of nodes and hubs, with the hubs being connected by random assortative or gradient connections\cite{BrajNeel}. In the case of the second network, in addition to nearest neighbour connections between nodes, the nodes are connected probabilistically to other nodes, with the probability of a connection between nodes being dependent on the Euclidean distance between them\cite{waxmangraph}. Such networks are called Waxman networks and are popular models of internet topology\cite{lakhina}. Synchronisation effects are observed in the congested phase of both these model networks. In addition to these two networks, we also discuss synchronisation effects seen in actual internet data.

We first study synchronization behavior in a two dimensional
communication network of nodes and hubs. Such networks
have been considered earlier in the context of search algorithms
\cite{kleinberg}
and of network traffic                                        
with routers and hosts \cite{Ohira,Sole2,fuks}. Despite the regular $2-d$ geometry such models have shown log-normal distribution in latency times as seen in Internet dynamics \cite{sole1}. The lattice consists of
two types of nodes, the regular or ordinary nodes, which are connected
to each of their nearest neighbors, and the hubs, which are connected to
all                                                           
the nodes in a given area of influence, and are randomly distributed in
the lattice.                                                  
Thus, the network represents a model with local clustering and
geographical separations \cite{warren,cohen}. Congestion effects are
seen on this network when a large number of messages travel 
between multiple sources and targets due to various factors   
like capacity, band-width and network                         
topology \cite{Huang}. Decongestion strategies,  which        
involve the                                                   
manipulation of factors                                       
like capacity and connectivity have been set up for these networks.
Effective connectivity strategies have focused on             
setting up random assortative\cite{braj1}, or gradient connections\cite{sat}
 between hubs of
high betweenness centrality.                                  
                              
We introduce the ideas of phase synchronization and
complete synchronization in the context of the queue lengths at the hubs. The queue at a given hub is defined to
be the number of messages which have the hub as a temporary target. During
multiple message transfer, when many messages run simultaneously on the
lattice, the network tends to congest when the number of messages exceed
a certain critical number, and the queue lengths tend to build up at
hubs which see heavy traffic. The hubs which see heavy traffic are ranked by the co-efficient of         
betweenness centrality ($CBC$), which is the fraction of messages    
which pass through a given hub. We focus on the top five hubs ranked by
CBC. Phase synchronization is seen between pairs of hubs of comparable betweenness centrality. The hub
which is slowest to decongest (generally the hub of  
 highest CBC) drives the slower hubs with a cascading  
master-slave effect in the hub hierarchy.       
When the network starts decongesting, the queue lengths decrease, and
synchronization is lost. These results are reflected in the global
synchronization parameter. When decongestion strategies which set up random assortative, or gradient, connections between hubs are implemented,   complete synchronization is seen between some pairs of these hubs in the congested phase, and phase synchronization is seen between others.
We demonstrate  our results  in the context
of the gradient decongestion strategy, but the results remain
unaltered for decongestion strategies based on random assortative
connections. 
Similar results are seen for constant density traffic where a fixed number of messages are fed on the system at regular intervals. Total synchronization is also seen in the queue lengths of the hubs of high CBC.

All the results obtained for the first model are observed for message transport on Waxman topology network, where again synchronisation of hubs of high CBC is observed in the congested state. 
We demonstrate these results.
Finally we study internet traffic data and demonstrate that phase synchronisation is seen in this data as well. Intermittent phase synchronization is also seen in this data.

\section{ A communication network with local clustering and geographic separation }                                    

\begin{figure*}                                               
\begin{center}
\begin{tabular}{cccc}
(a)&
\includegraphics[width=7.5cm,height=7cm]{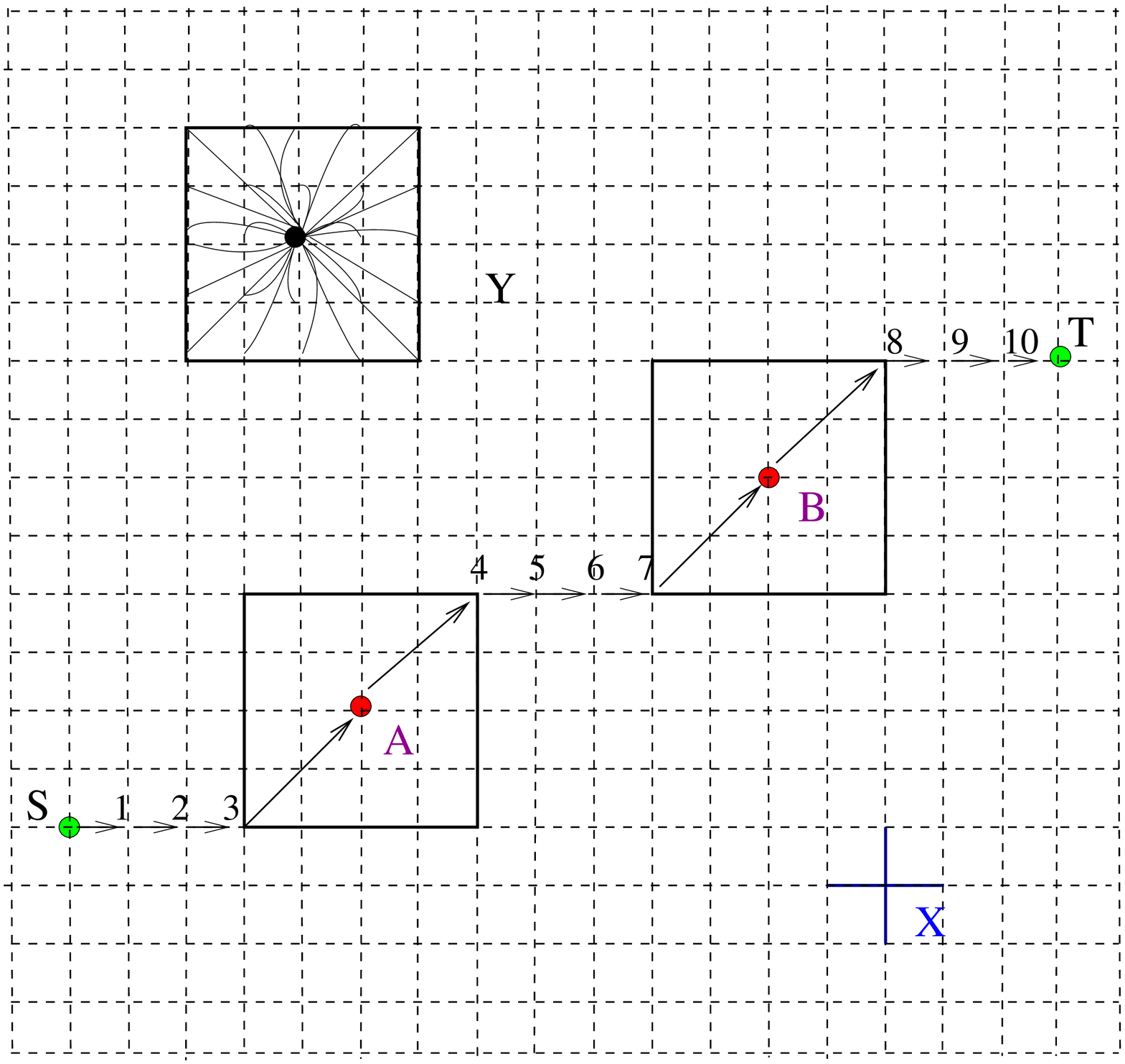}&   
\hspace{1.5cm}                                          
(b)&
\includegraphics[width=7.5cm,height=7cm]{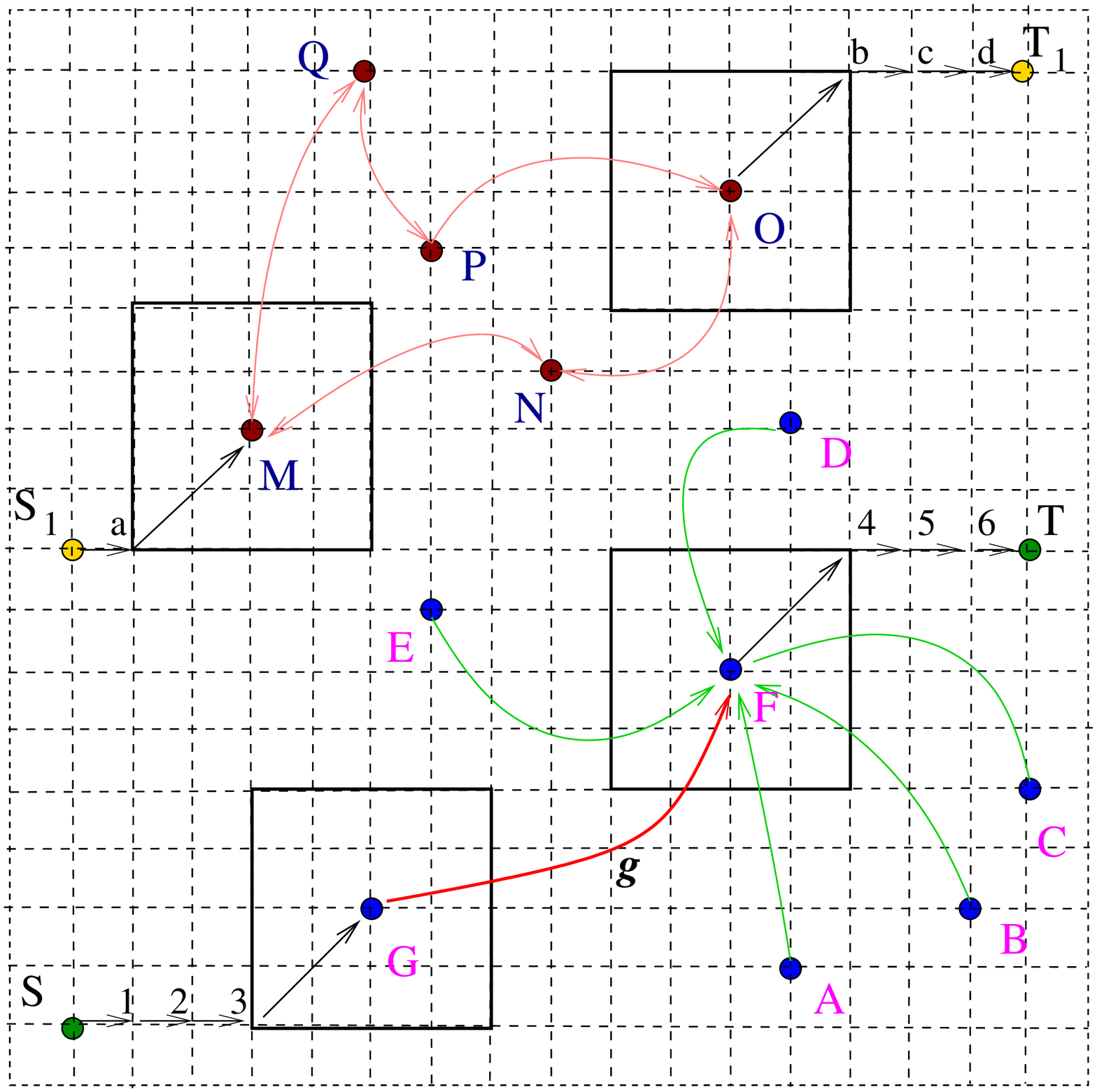}\\
\end{tabular}  
\end{center}
\caption{\label{fig:asgr}(Color online) (a) A regular two dimensional lattice. The node {\it X} is an ordinary node with nearest neighbor connections. Each hub has a square influence region (as shown for the hub {\it Y}). A typical path from the source $S$ to the target $T$ is shown with labeled sites. The path $S-1-2-3-$A$-4-5-6-7-$B$-8-9-10-T$ passes through the hubs {\it A} and {\it B}. (b) The high CBC hubs (M-Q) are connected by the $CBC_{a}$ mechanism. A message is routed along the
shortest path $S_{1}-a-$M$-$N$-$O$-b-c-d-T_{1}$. We enhance the capacities of high CBC hubs (A-G) proportional to their CBC values by a factor of 10. Connections between these hubs are made by the gradient mechanism. After the implementation of the gradient mechanism, the distance between $G$ and $F$ is covered in one step as shown by the link $g$ and a  message is routed along the path $S-1-2-3-$G$-$g$-$F$-4-5-6-T$.\\}
\end{figure*}

We first study traffic congestion for                               
a                                                             
model network with local clustering developed in Ref.\cite{BrajNeel}.
This network consists of a two{-}dimensional lattice with ordinary nodes and hubs (See Fig. \ref{fig:asgr}). Each ordinary node is connected to its nearest-neighbors, whereas                
the hubs are connected to                                     
all nodes  within                                             
a given  area of influence                                    
defined as a square of side $2k$ centered around the hub\cite{BrajNeel}. The hubs are randomly distributed on the lattice              
such that no two hubs are separated by less than a minimum distance,
$d_{min}$. Constituent nodes in the overlap areas of hubs  acquire
connections to all the hubs whose influence areas overlap. The source S$(is,js)$ and target T$(it,jt)$ are chosen from the  lattice and separated by a fixed distance $D_{st}$ which is defined by the Manhattan distance $D_{st}$ = $|is-it|$ + $|js-jt|$ . 
It is useful to identify and rank hubs which see the maximum traffic.
This is done by defining the co-efficient of betweenness centrality
(CBC) where  the CBC of a given hub $k$ is defined as
$CBC=\frac{N_{k}}{N}$, i.e.
the ratio of the number of messages  that go through a hub $k$ to the
total number of messages running on the lattice. These are listed in
Table~\ref{tab:table1}.

Efficient decongestion
strategies have been set up by connecting hubs of high CBC amongst
themselves, or to randomly chosen other hubs via assortative
connections
\cite{braj1}. Gradient mechanisms \cite{grad} can also be used to
decongest traffic \cite{danila, sat}(See Fig. \ref{fig:asgr}(b)). 

\begin{table*}                                                
\caption{\label{tab:table1}This table shows the CBC values and ranking of
the top five hubs. A total number of 2000 messages are traveling
simultaneously on a ${100\times 100}$ lattice with 4$\%$ hub density and
$D_{st}$ = 142 and run time set at 5000.}
\begin{ruledtabular}                                          
\begin{tabular}{ccc}                                          
Hub label & CBC value & Rank\\                                
\hline                                                        
x &  0.827  & 1 \\                                            
y &  0.734  & 2 \\                                            
z &  0.726  & 3 \\                                            
u &  0.707  & 4 \\                                            
v &  0.705  & 5\\                                             
\hline                                                        
\end{tabular}
\end{ruledtabular}
\end{table*}

In all the simulations here,  we consider a lattice of size ${100\times
100}$ with 4$\%$ hub density and $D_{st}$ = 142, $d_{min}=1$. The critical message density which congests this lattice is $N_c=1530$. The studies carried out here correspond to the congested phase, where $2000$
or $4000$ messages run on the lattice. We first consider the baseline lattice as in Fig. \ref{fig:asgr}(a) where there are no
short-cuts between the hubs. The message holding capacity of ordinary nodes and hubs is unity for the baseline lattice.
                                                              
\begin{figure*}                                               
\begin{center}
\begin{tabular}{cccc}  
(a)&              
\includegraphics[width=7.5cm,height=7cm]{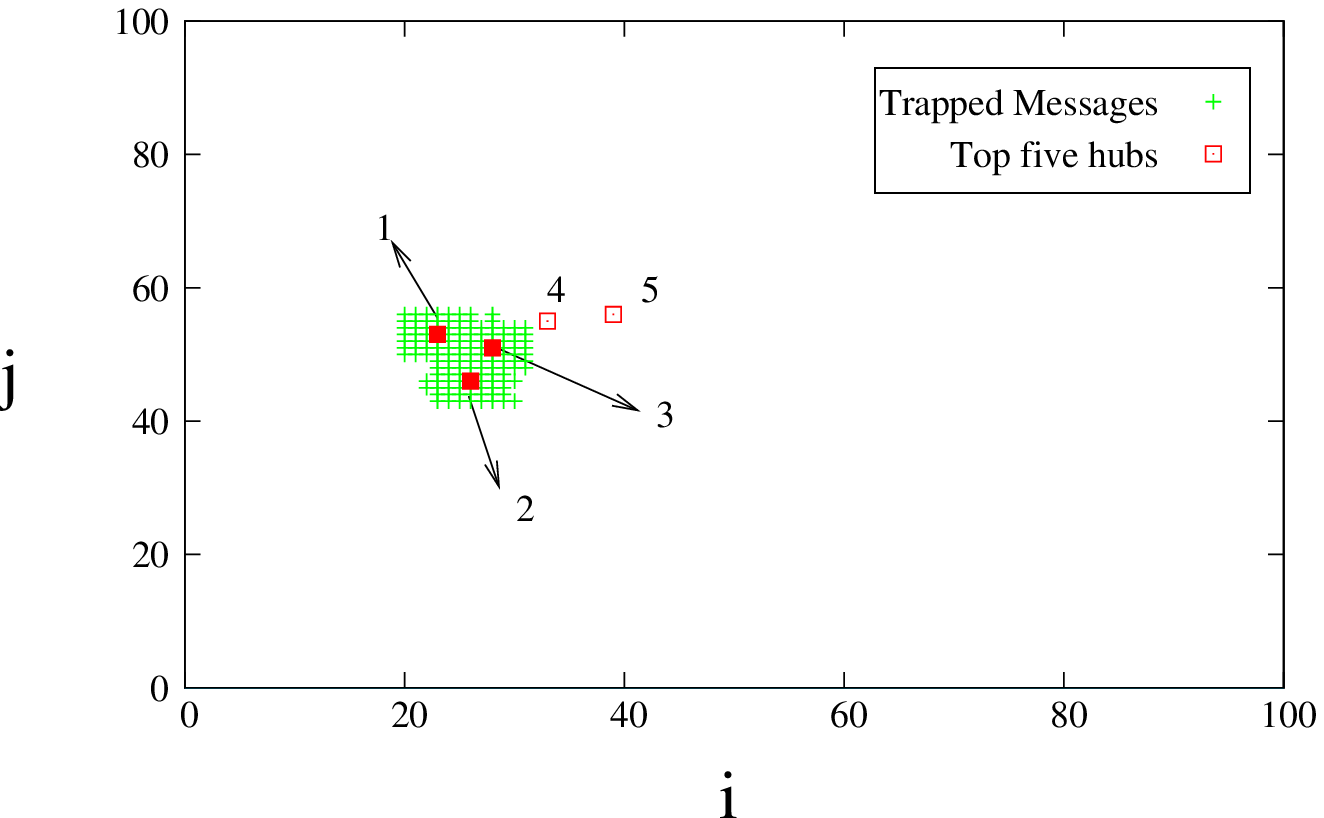}&   
\hspace{1.5cm} 
(b)&
\includegraphics[width=7.5cm,height=7cm]{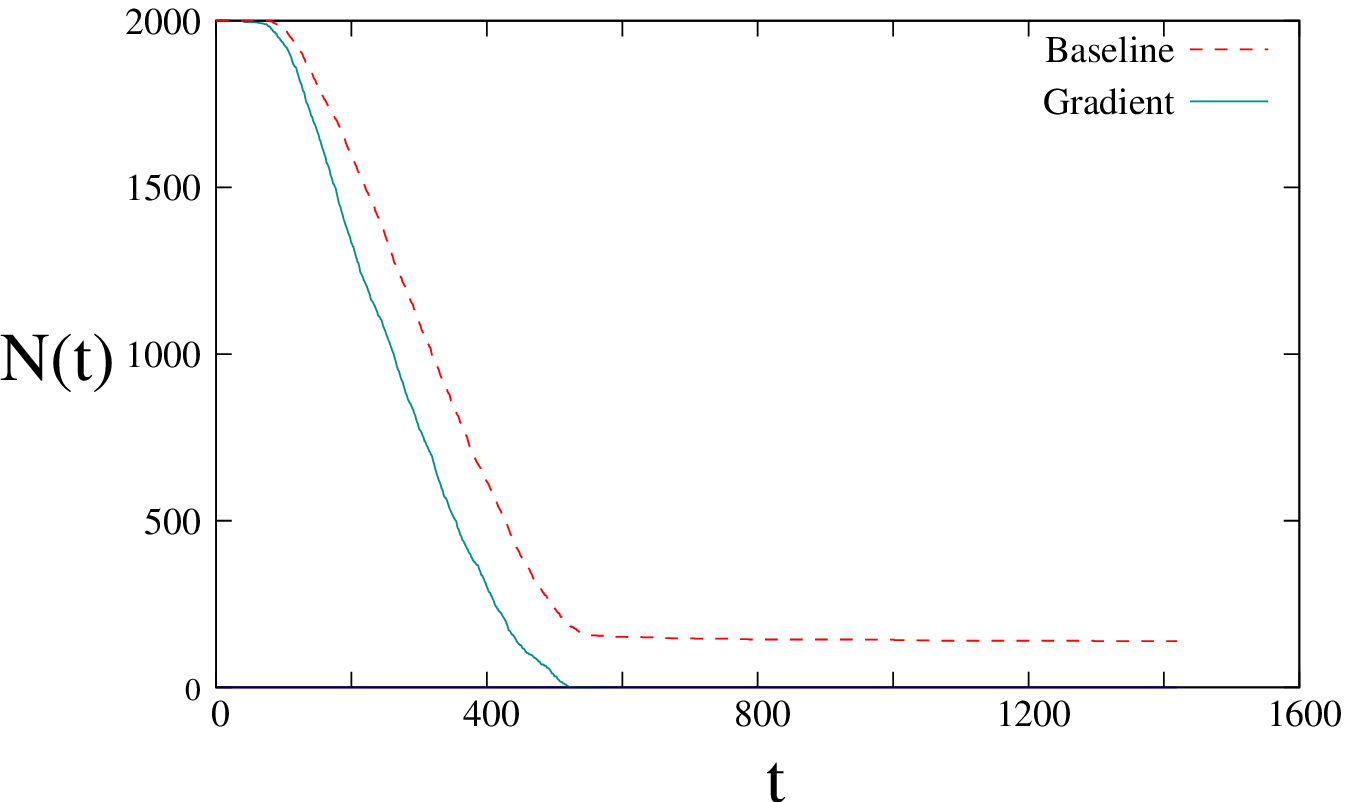}\\
\end{tabular}  
\end{center}
\caption{\label{fig:trap1}(Color online) (a) Figure shows messages trapped in the vicinity of top five hubs ranked by their CBC values. Numbers indicate the ranks of the top five hubs. A total number of 133 messages are trapped in the lattice when 2000 messages are traveling simultaneously on a $100\times 100$ lattice with 4\% hub density for a run time of 10$D_{st}$. (b) Messages are trapped in the lattice when the hubs are not connected (baseline mechanism). All messages are delivered to their respective targets once the top five hubs ranked by their CBC values are connected by the gradient mechanism. \\}
\end{figure*}

A given number $N$ of source and target pairs separated by    
a fixed distance $D_{st}$ are                                 
randomly selected on the lattice. Here, all source nodes start sending
messages to the selected recipient
nodes simultaneously, however, each node can act as a source for
only one message during a given run. 
The routing takes place by a distance based algorithm in which  
each node holding a message directed towards a target tries to identify
the hub nearest
to itself, and                                                
in the direction of the target as the temporary target, and tries to
send the message to the temporary target through the connections
available to it. During peak traffic, when many messages run, some of the
hubs, which
are located such that many paths pass through them, have to   
handle more messages than they are capable of holding simultaneously.       
Messages tend to jam in the vicinity of such hubs (usually the hubs of
high CBC) leading to formation of transport traps which leads to congestion in the network. Other factors like the opposing movement of messages from sources and targets situated on different sides of the lattice, as well as edge effects ultimately result in the formation of transport traps. We have studied trapping configurations for the same $2-d$ network in Ref.\cite{sat}. Fig. \ref{fig:trap1}(a) shows a situation in which messages are trapped in the vicinity of high CBC hubs. Fig. \ref{fig:trap1}(b) shows the number of messages running on the lattice as a function of time. It is clear that the messages are trapped for the baseline case. We study the network for situations which show this congested phase.
                                                              
\section{Queue lengths and synchronization}                   
                                                              
\begin{figure*}                                               
\begin{center}                                                
\begin{tabular}{cccc}    
(a)&                                       
\includegraphics[width=7.5cm,height=7cm]{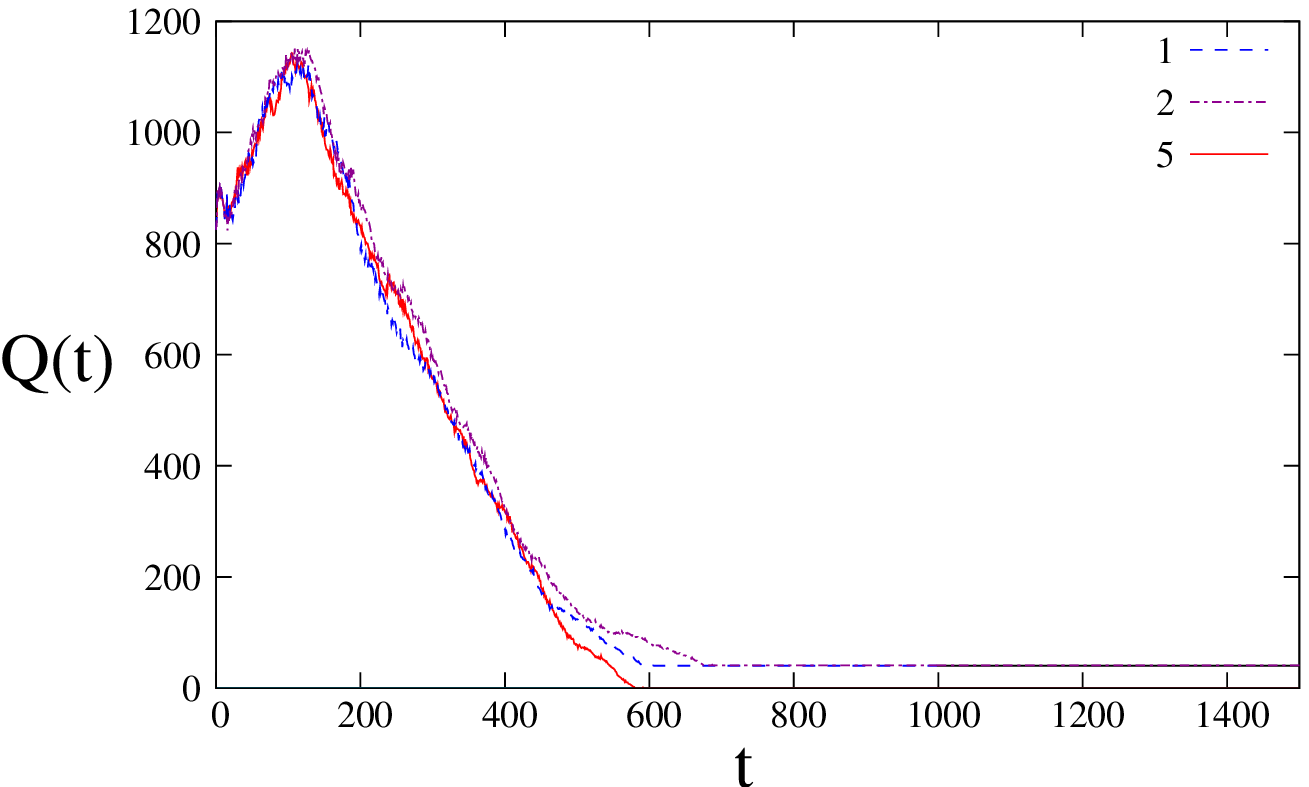}&    
\hspace{1.5cm}  
(b)&
\vspace{0.3cm}
\includegraphics[width=7.5cm,height=7cm]{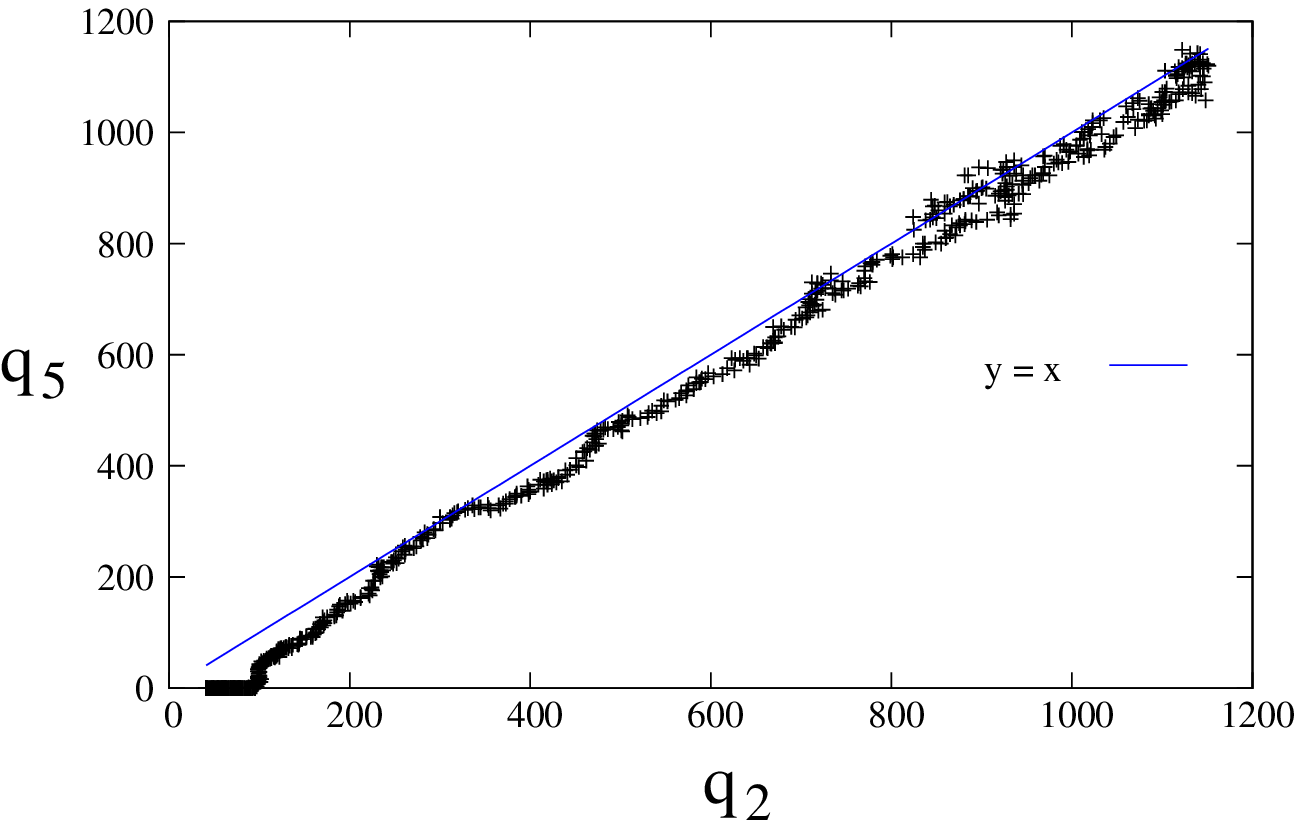}\\ 
(c)&
\includegraphics[width=7.5cm,height=7cm]{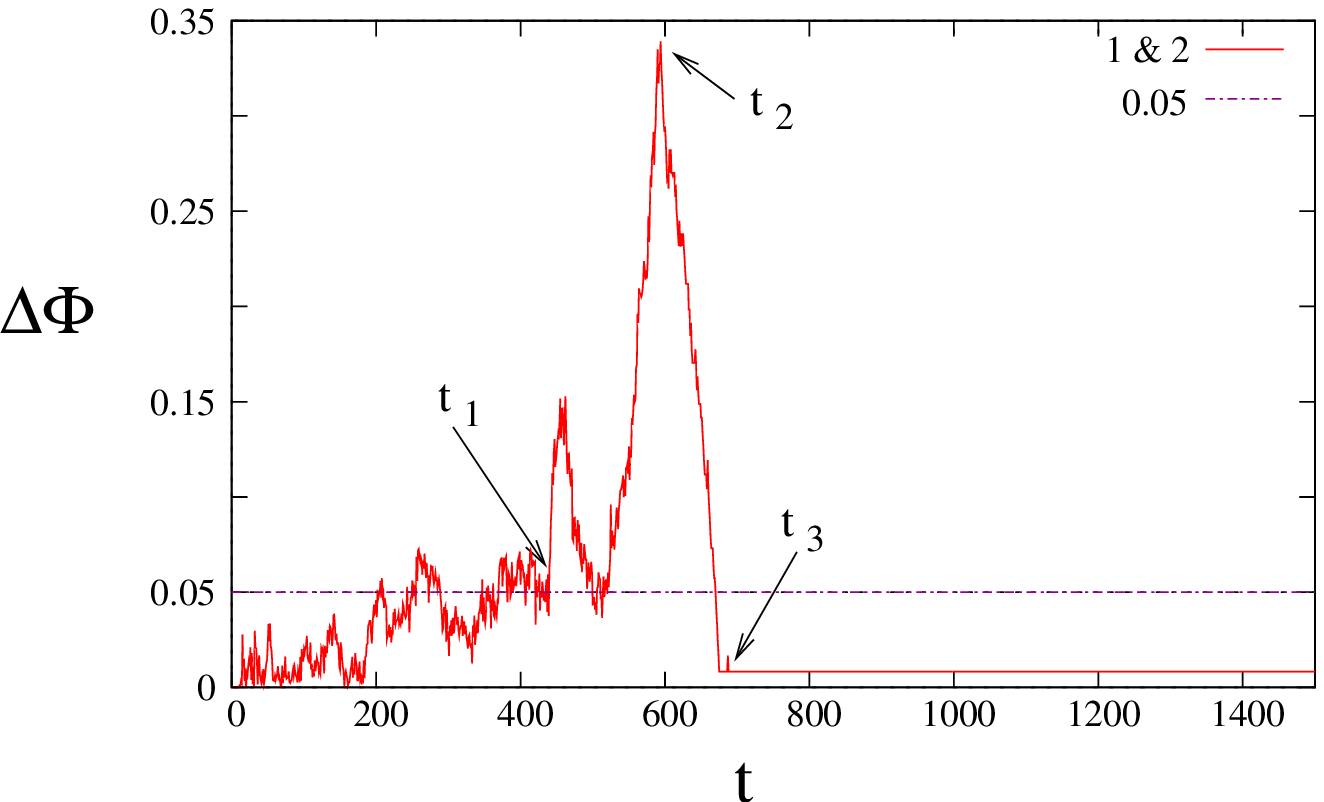}&  
\hspace{1.5cm}    
(d)&
\includegraphics[width=7.5cm,height=7cm]{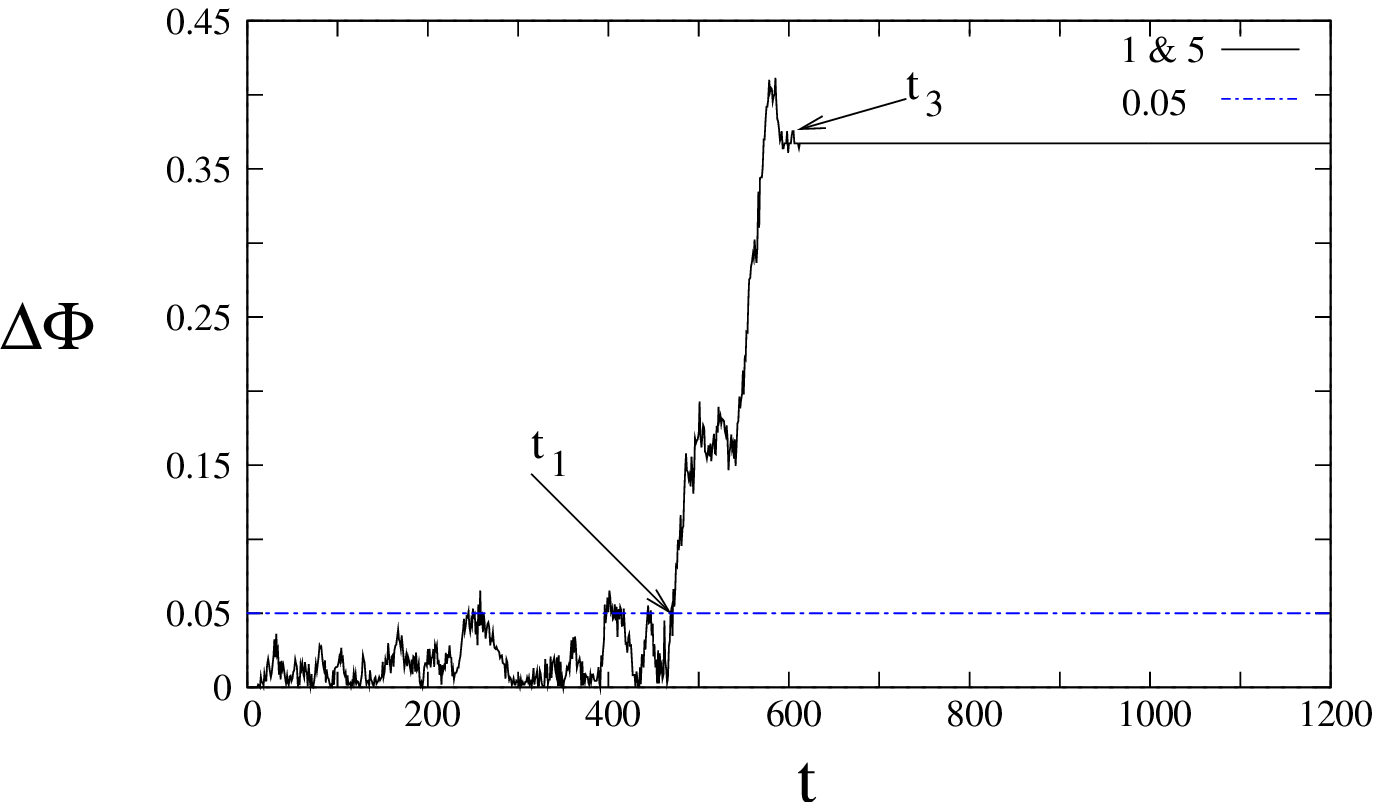}\\   
\end{tabular}                                                 
\end{center}                                                  
\caption{\label{fig:psmodel1}(Color online)We consider the baseline mechanism. (a) Queue lengths for the
hubs ranked $1$, $2$ and $5$ ranked by $CBC$. (b) Complete synchronization is not observed in the queue lengths. The error in the fit to $y=x$ is 1.940. (c) Phase synchronisation of the queue lengths of the two top most hubs. (d) Phase synchronization of the queue lengths of the $1_{st}$ and $5_{th}$ hub ranked by CBC. A total number of 2000 messages are traveling simultaneously on a $100\times 100$ lattice with 4\% hub density.\\}
\end{figure*}
                                                              
As mentioned in the introduction, the queue at a given hub is defined to
be the
number of messages which have the hub as a temporary target.  
As traffic increases in the network, hubs which see heavy traffic
start getting choked due to capacity limitations, and are unable to         
transfer  messages aimed towards them to the next temporary target.         
Thus  queue lengths start to build up at                      
these hubs. If these hubs are not decongested quickly, so that the queue    
lengths start falling, the congestion starts spilling over to other         
hubs. If the number of messages increases beyond a certain critical         
number, messages get trapped irretrievably and the entire lattice
congests. A plot of the queue lengths   as a function of time can be
seen in Fig. \ref{fig:psmodel1}(a). Here, the queues at the $1_{st}$, $2_{nd}$ and $5_{th}$ hubs ranked by the CBC are
plotted for the base-line network with no decongestion strategies implemented.
Thus the network congests very easily. Since the queue                                               
length is defined as the number of messages with the given hub as the       
temporary target, the queue starts dropping as soon as the hub starts       
clearing messages and reaches a minimum.                      
Meanwhile, other hubs which were                              
temporary targets have cleared their messages, and some new messages        
pick up the hub of interest as their temporary target. The queue thus       
starts building up here, and reaches a maximum. After this the  messages start clearing, and the queues       
drop sharply. However, since the number of messages is sufficiently large    
for the network to congest, some messages get trapped in the vicinity of    
the hub, and the queues saturate to a constant value.                                                
Similar phenomena can be seen at the hubs of lower CBC (see Fig. \ref{fig:psmodel1}(a)). Here again three distinct
scales can be seen with values of the same order as those for the
highest ranked hub. An important difference can be seen in the queues
of the fifth ranked hub (Fig. \ref{fig:psmodel1}(a)) as well as the fourth ranked hub (not shown). Since these hubs have lower CBC
values, and thus fewer messages take them as the temporary targets, the
queues at these hubs clear completely. Thus, the saturation value at
these hubs, is zero. It should also be noted that the time at which
the last two hubs clear completely, i.e. the queue length drops to zero,
is substantially earlier  than the saturation time
of the top two hubs. The above results are observed for a typical configuration and is valid for different configurations as well.

\subsection{Synchronization}

We now study the synchronization between the queues at different hubs.
We see phase synchronization between queues at pairs of high CBC hubs
for the baseline, and complete synchronization between some
 pairs once decongestion 
strategies are implemented.
The usual definitions of complete
synchronization and phase synchronization in the literature are as
follows.

Complete synchronization (CS) in coupled identical systems appears as
the equality of the state variables while evolving in time. Other names
were given in the literature, such as {\it conventional synchronization}
or {\it identical synchronization}\cite{pyragas}. It has been observed that for
chaotic oscillators starting from uncoupled non-synchronized oscillatory
systems, with the increase of coupling strength, a weak degree of
synchronization, the {\it phase synchronization}(PS) where the phases
become locked is seen \cite{pikovsky,rosa} . Classically, the phase synchronization of
coupled periodic oscillators is defined
as the locking of phases ${\phi_{1,2}}$ with a ratio $n:m$ ($n$ and $m$
are
integers), i.e. $|n\phi_{1}-m\phi_{2}|< $Const.

These two concepts of                                         
synchronization are applied to the queue lengths of the top five hubs. The plot of $q_{i}(t)$ as a function of average queue length $<q(t)>$ shows a loop in the congested phase, similar to that observed in coupled chaotic oscillators \cite{maria}. We define a phase as in \cite{maria}  ${\Phi}_{i}(t)=tan^{-1}(\frac{q_{i}(t)}{<q(t)>})$, where $q_{i}(t)$ is the queue length of $i_{th}$ hub at time $t$, and $<q(t)>=\frac{1}{N_{h}}\displaystyle\sum_{i}q_{i}(t)$ where the average is calculated over the top five hubs ($N_{h}=5$). The queue lengths are phase synchronized if                                         
\begin{equation}                                              
|\Phi_{i}(t)-\Phi_{j}(t)| < Const                             
\end{equation}                                                
where $\Phi_{i}$(t) and $\Phi_{j}$(t) are the phase at time $t$ of the      
$i_{th}$ and $j_{th}$ hub respectively.                       
                                                              
Two queue lengths $q_{i}(t)$ and $q_{j}(t)$ are said to be completely       
synchronized if                                               
\begin{equation}                                              
q_{i}(t)=q_{j}(t)                                             
\end{equation}                                                
                                                              
Fig. \ref{fig:psmodel1}(b) shows that the queue lengths of the first and fifth ranked hubs are not completely synchronized. Fig. \ref{fig:psmodel1}(c) shows the phase difference between the top pair of hubs as a      
function of time for the base-line case. It is clear that     
the two hubs are phase synchronized in the regimes where the queues
congest. There are three distinct time scales in the problem. 
The two hubs are phase synchronized up to the first time scale $t_1$,
where 
the queues cross each other first, they lose synchronization after this.
The point
at which the phase difference is maximum is $t_2$.            
This is the point at which the first hub saturates, but the second hub      
is still capable of clearing its queue.                       
At $t_3$ both the hubs get                                    
trapped and the phases lock again.

Fig. \ref{fig:psmodel1}(d) shows a similar plot for the hubs of the two        
remaining ranks. It's clear that the hubs phase synchronize. The synchronization behavior of the remaining hubs for a typical configuration is listed in Table~\ref{tab:table2}. It is clear that the hubs synchronize pair wise, and that the slower hubs drive the hubs which clear faster. Since the queues at the fourth and fifth hub clear faster than the first hub saturates, there is no peak in the $\Delta \Phi$\footnote{$\Delta \Phi$ = $|\Phi_{i}(t)-\Phi_{j}(t)|$} plot for the $(1,5)$ pair and hence no scale $t_2$. The phase synchronization between hubs three and four shows similar behavior. This is valid for different configurations as well.                                                               
                                                              
\begin{table*}                                                
\caption{\label{tab:table2}The table shows the pair of queue lengths of     
the top five hubs which are phase synchronized for a typical configuration for the baseline
mechanism. All
the pairs of top five hubs are phase synchronized within a constant
$C=0.05$. The three time scales $t_{1}$, $t_{2}$ and $t_{3}$ for every      
synchronized pair are shown in the table below. Similar results are observed for different configurations as well. We consider a ${100\times 100}$ lattice with 4$\%$ hub density and $D_{st}$ = 142 . A total number of 2000 messages are traveling simultaneously in the lattice. The run time is set at $5000$.}                      
\begin{ruledtabular}                                          
\begin{tabular}{cccc}                                         
PS pairs & $t_{1}$ & $t_{2}$ & $t_{3}$ \\                     
\hline                                                        
$(1,2)$ & 440 & 589 & 675 \\                                  
$(1,3)$ & 225 & 595 & 727 \\                                  
$(1,4)$ & 360 & 595 & 720 \\                                  
$(1,5)$ & 472 & - & 590 \\                                    
$(2,3)$ & 295 & 495 & 727 \\                                  
$(2,4)$ & 405 & 620 & 727 \\                                  
$(2,5)$ & 450 & 590 & 675 \\                                  
$(3,4)$ & 285 & - & 727 \\
$(3,5)$ & 270 & 585 & 727 \\                                  
$(4,5)$ & 360 & 585 & 727 \\                                  

\end{tabular}                                                 
\end{ruledtabular}
\end{table*}

\begin{figure*}                                               
\begin{center}
\includegraphics[width=7.5cm,height=7cm]{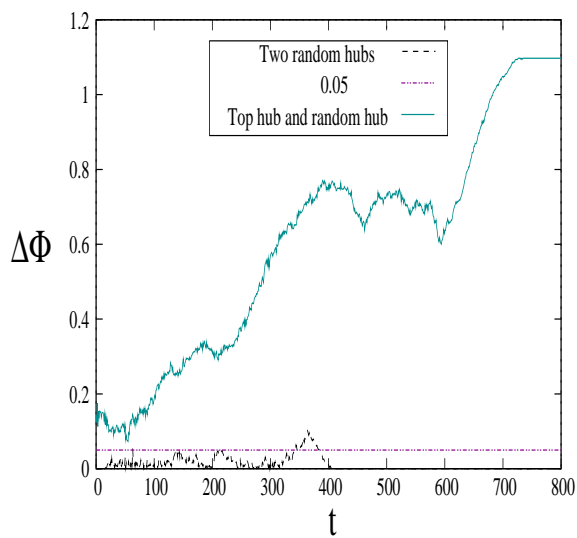}\\   
\end{center}
\caption{\label{fig:randhub}(Color online) No phase synchronization is observed between top most hub and a randomly chosen hub. The two randomly chosen hubs are phase synchronized to large extent. All parameters as in Fig. \ref{fig:psmodel1}. A total number of 2000 messages are traveling simultaneously on a $100\times 100$ lattice with 4\% hub density.\\}
\end{figure*}             

It is also interesting to compare the synchronization effects between
these hubs of high CBC, and randomly selected  hubs on the lattice. Fig. \ref{fig:randhub}
shows the phase difference between the hub of highest CBC (hub `x',
ranked 1) and a randomly chosen hub (with CBC value $0.56$ ). It is clear that there is
no synchronization between these two hubs. However, this randomly chosen
hub shows excellent phase synchronization with another randomly chosen hub (with comparable CBC value $0.67$). Similar results are seen for larger number of messages.

\subsection{Decongestion strategies and the role of connections}                          

\begin{figure*}                          
\begin{center}
\includegraphics[width=7.5cm,height=7cm]{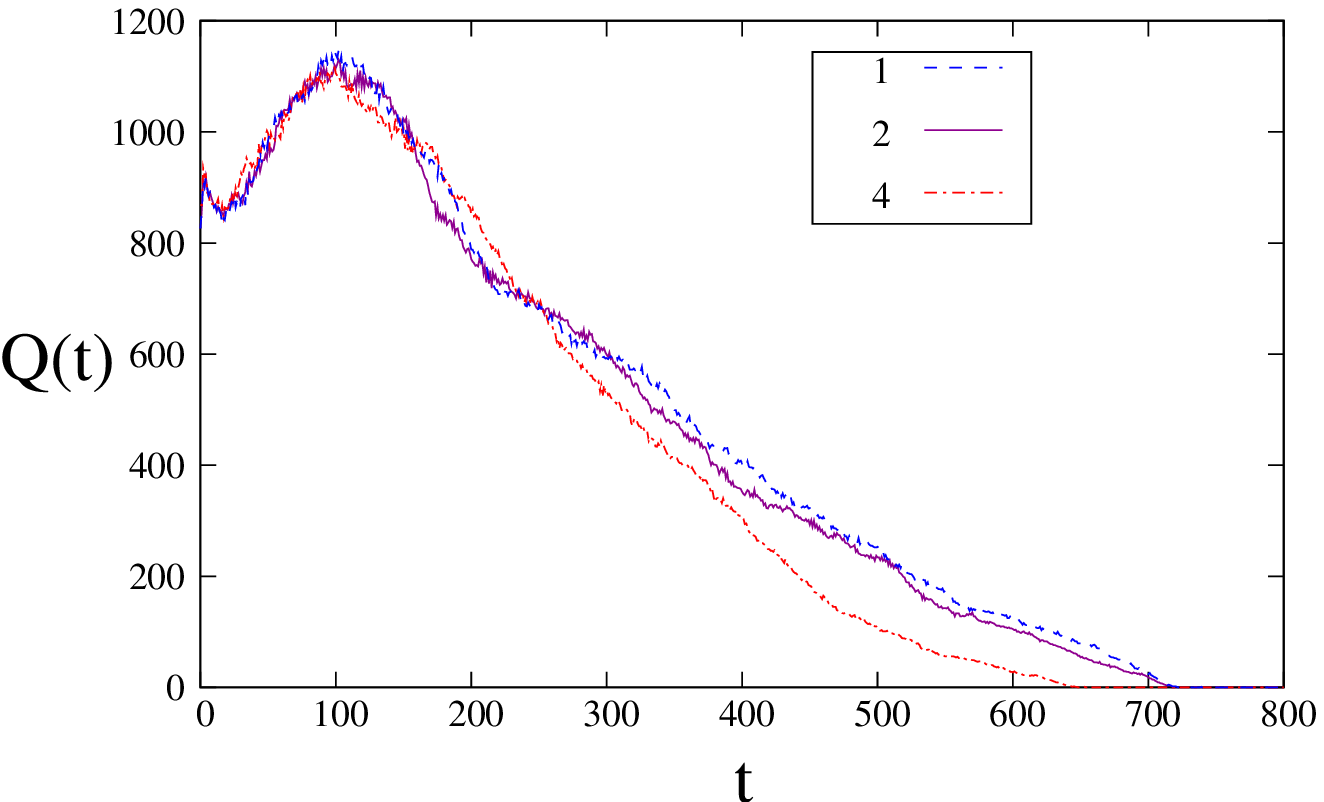}
\end{center}
\caption{\label{fig:bslgrad}(Color online) Plot of queue lengths of the $1_{st}$, $2_{nd}$ and $4_{th}$ hubs for the gradient mechanism. A total number of 2000 messages are traveling simultaneously on a $100\times 100$ lattice with 4\% hub density.\\}
\end{figure*}             

\begin{figure*}
\begin{center}
\begin{tabular}{cccc}
(a)&
\includegraphics[width=7.5cm,height=7cm]{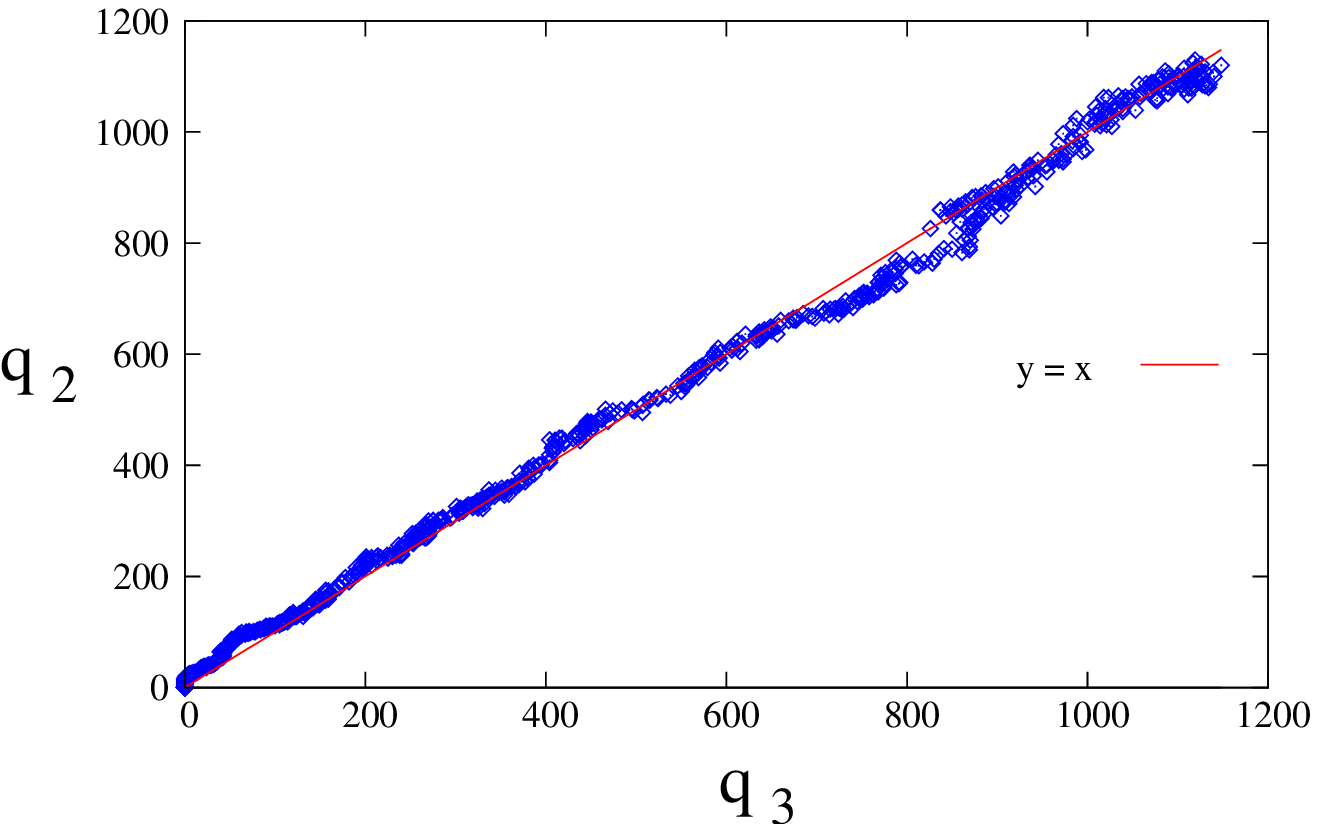}&
\hspace{1.5cm}  
(b)&
\includegraphics[width=7.5cm,height=7cm]{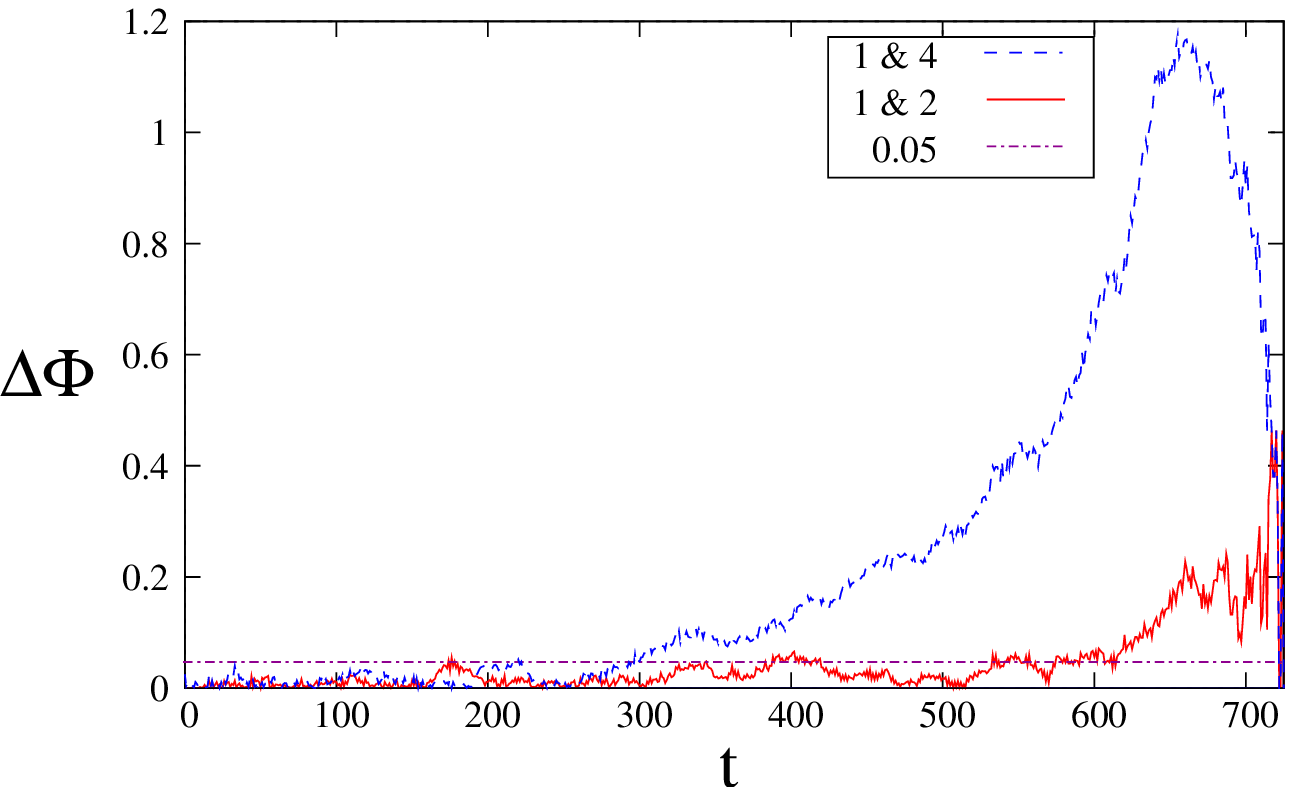}\\
\end{tabular}
\end{center}
\caption{\label{fig:csps1}(Color online) Figure shows (a) complete synchronization in queue
lengths for the $2_{nd}$ and $3_{rd}$ hub and (b) phase synchronization in queue lengths for pair(s) of top five hubs when connected by the $gradient$ mechanism. A total number of 2000 messages are traveling simultaneously on a $100\times 100$ lattice with 4\% hub density. The error in the fit $y=x$ in (a) is 0.992.\\}
\end{figure*}

 As discussed earlier, the addition of extra connections between the hubs
of high betweenness centrality can ease congestion. This leads to two       
effects. The time scales of the problem, the rate at which the queues       
build up and clear, and the way in which correlations occur between
different hubs are altered due to the addition of extra connections.        
We see the effects of this in the  synchronization between the queues at    
the hubs. We illustrate the effects seen for gradient connections  between 
the hubs (Fig. \ref{fig:asgr}(b)). To set up the gradient mechanism, we enhance the capacities of top five hubs ranked by their CBC values, proportional to their CBC values by a factor of 10. A gradient flow is assigned from each hub to all the hubs with the maximum capacity ($C_{max}$). Thus, the hubs with lower capacities are connected to the hubs with highest capacity $C_{max}$ by the gradient mechanism. Hence the hub with highest CBC value is maximally connected. Fig. \ref{fig:trap1}(b)  and Fig. \ref{fig:bslgrad} shows that connecting the top five hubs by the gradient mechanism relieves the system of congestion rapidly when 2000 messages are traveling in the lattice for 4\% hub density and run time of $10D_{st}$.

The most striking observation is that now complete synchronization is seen
between at least one pair of hubs, and phase synchronization is seen
between the remaining pairs. In Fig. \ref{fig:csps1}(a) we plot a pair of
queue length $q_{i}$ vs $q_{j}$.     
If these two quantities lie along the $y=x$ line with a standard deviation  less than one, we call them completely     
synchronized. It is clear from the Fig. \ref{fig:csps1}(a) and the value of the standard deviation that the queue lengths of the $2_{nd}$ and the $3_{rd}$ ranked hubs are completely synchronized. These two hubs are of comparable CBC values (See Table~\ref{tab:table1}) and are indirectly connected via the top most ranked hub, to which each of the lower ranked hubs is connected via a gradient. If the standard deviation is greater than one, the queue lengths are not completely synchronized. In Fig. \ref{fig:csps1}(b) we observe phase synchronization when the top five hubs are connected by the gradient mechanism. Phase synchronization is observed when the queues congest. As soon as the queues decongest they are no longer phase synchronized. This observation is true for the complete synchronization as well.  In the gradient scheme we see a star-like geometry where the central hub is connected to the hubs of low capacity. This central hub gets congested leading to the congestion of 
 the rest of the hubs. Once this hub gets decongested the rest of the hubs of high CBC get cleared. Thus, the central
hub, which is the hub of highest CBC, drives the rest.         

\subsection{The finite time Lyapunov exponent}

\begin{figure*}
\begin{center}
\begin{tabular}{cccc}
(a)&
\includegraphics[width=7.5cm,height=7cm]{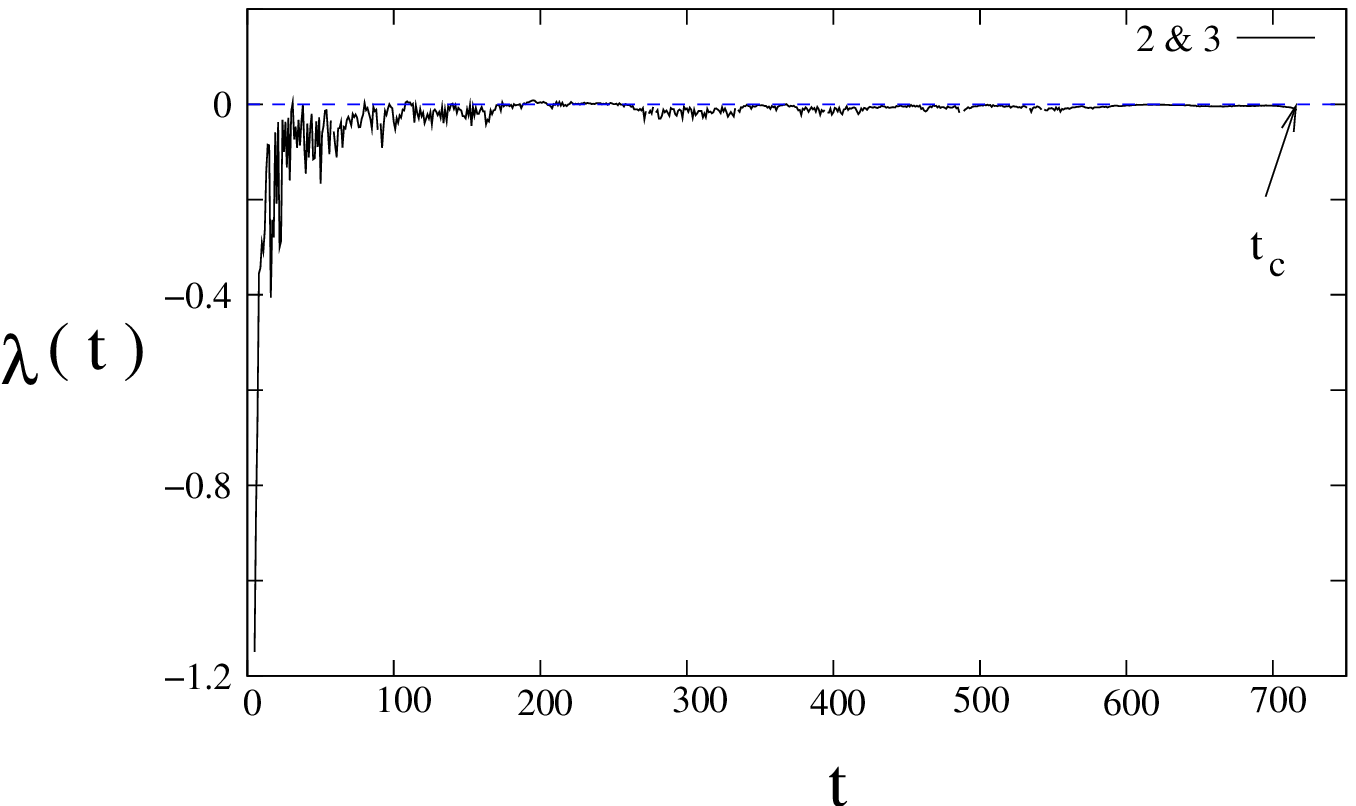}&
\hspace{1.5cm}  
(b)&
\includegraphics[width=7.5cm,height=7cm]{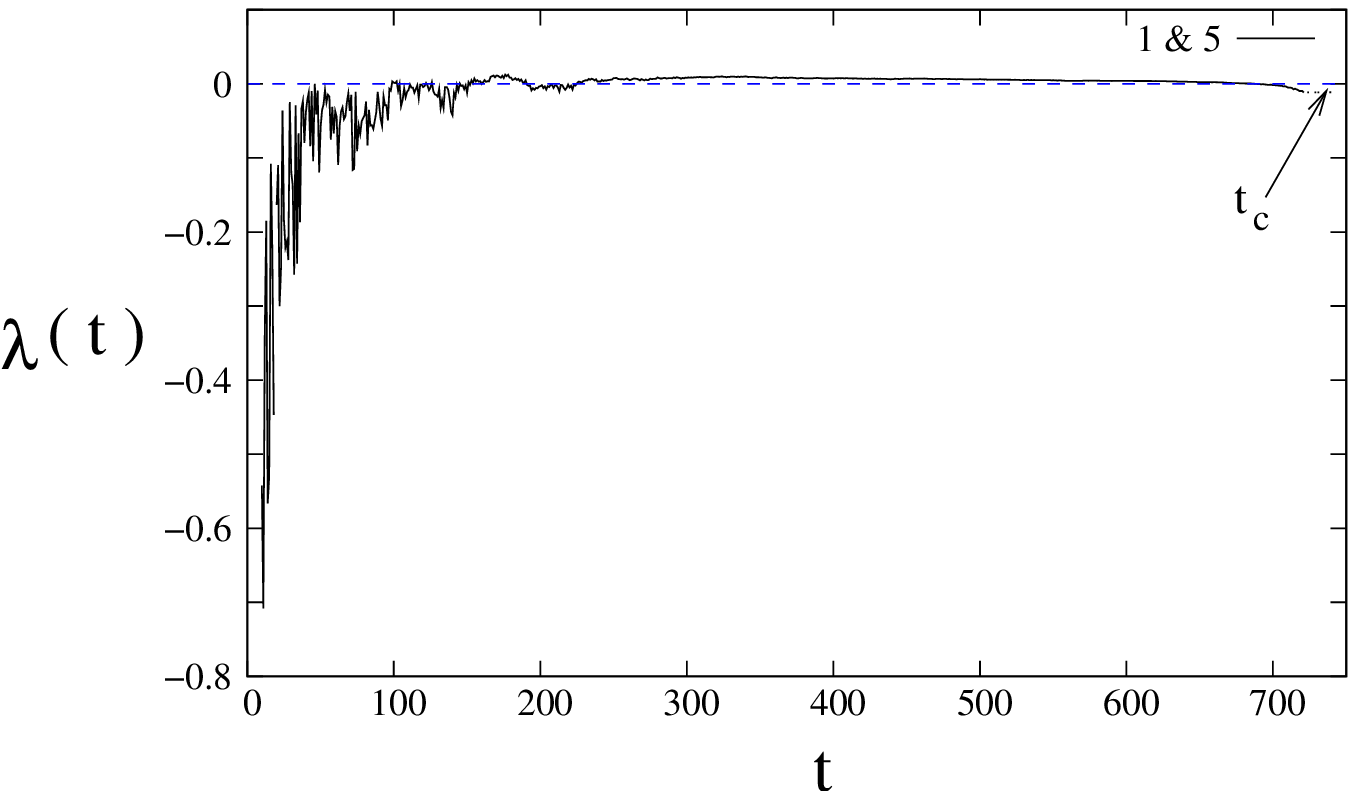}\\
\end{tabular}
\end{center}
\caption{\label{fig:lyap1}(Color online) Plot of the Lyapunov exponent for (a) complete synchronization in queue
lengths of $2_{nd}$ and $3_{rd}$ hub and (b) No complete synchronization in queue lengths of $1_{st}$ and $5_{th}$ hubs when connected by the $gradient$ mechanism. $t_{c}$ is the time at which all the messages are delivered to the respective targets. In (a) $t_{c}=720$ and (b) $t_{c}=740$. A total number of 2000 messages are traveling simultaneously on a $100\times 100$ lattice with 4\% hub density.\\}
\end{figure*}

 The queue lengths increase in the congested phase and the difference between two queue lengths (i.e. queue lengths at distinct hubs) is small in this phase, as compared to the decongested phase, where the difference between queue lengths is large. This is analogous to the behaviour of trajectories in the chaotic regime where the separation between two co-evolving trajectories with neighbouring initial conditions increases rapidly,  as compared to the separation  in the periodic regime where it rapidly decreases. Hence the stability of the completely synchronized state seen in the gradient case can studied by calculating the finite time Lyapunov exponent of the separation of queue lengths for the  top five pairs of hubs. The finite time Lyapunov exponent is given by

\begin{equation}
\lambda(t)=\frac{1}{t}\ln(\frac{\delta(t)}{\delta(0)})
\end{equation}

where $\delta(t)$ = $|q_{i}(t)-q_{j}(t)|$ and $\delta(0)$ is the initial difference in queue lengths \footnote{The queue lengths $q(t)$ are of the order $10^2$ in the congested phase, whereas the $\delta(t)-s$ are of order $1$. Thus $\delta(t) << q(t)$.}. If $\lambda(t) < 0$ then queue lengths are completely synchronized (Fig. \ref{fig:lyap1}(a)) and if $\lambda(t) > 0$ then queue lengths are not completely synchronized (Fig. \ref{fig:lyap1}(b)). The time is calculated from the time ($t=15$) at which the queue starts building up in the lattice. It is clear from the Fig. \ref{fig:lyap1}(a) that complete synchronization exists till $t_{c}=720$. This is the time at which queues are cleared. In Fig. \ref{fig:lyap1}(b) complete synchronization exists till $t_{cs}=150$, when queues are building up in the lattice. No complete synchronization is observed after this, but queues are cleared at $t_{c}=740$ .

\subsection{Global synchronization}                              
It is useful to define an over-all characterizer of emerging collective behavior. The usual characterizer  of global synchronization is the order
parameter \cite{Arenas} defined by
                                                              
\begin{equation}                                              
r\exp{i\psi}=\frac{1}{N_{h}}\sum_{j=1}^{N} \exp{i \Phi_{j}}       
\end{equation}                                                
$N_{h}$=5, where we consider the top five hubs.                                                          
Here $\psi$ represents the average phase of the system, and the
$\Phi_j$-s  
are the phases defined in Eq. 1.                     
Here the                                                      
parameter $0 \leq r \leq 1$ represents the order parameter of the system    
with the value $r=1$ being the indicator of total synchronization.

\begin{figure*}
\begin{center}
\begin{tabular}{cccc}
(a)&
\includegraphics[width=7.5cm,height=7cm]{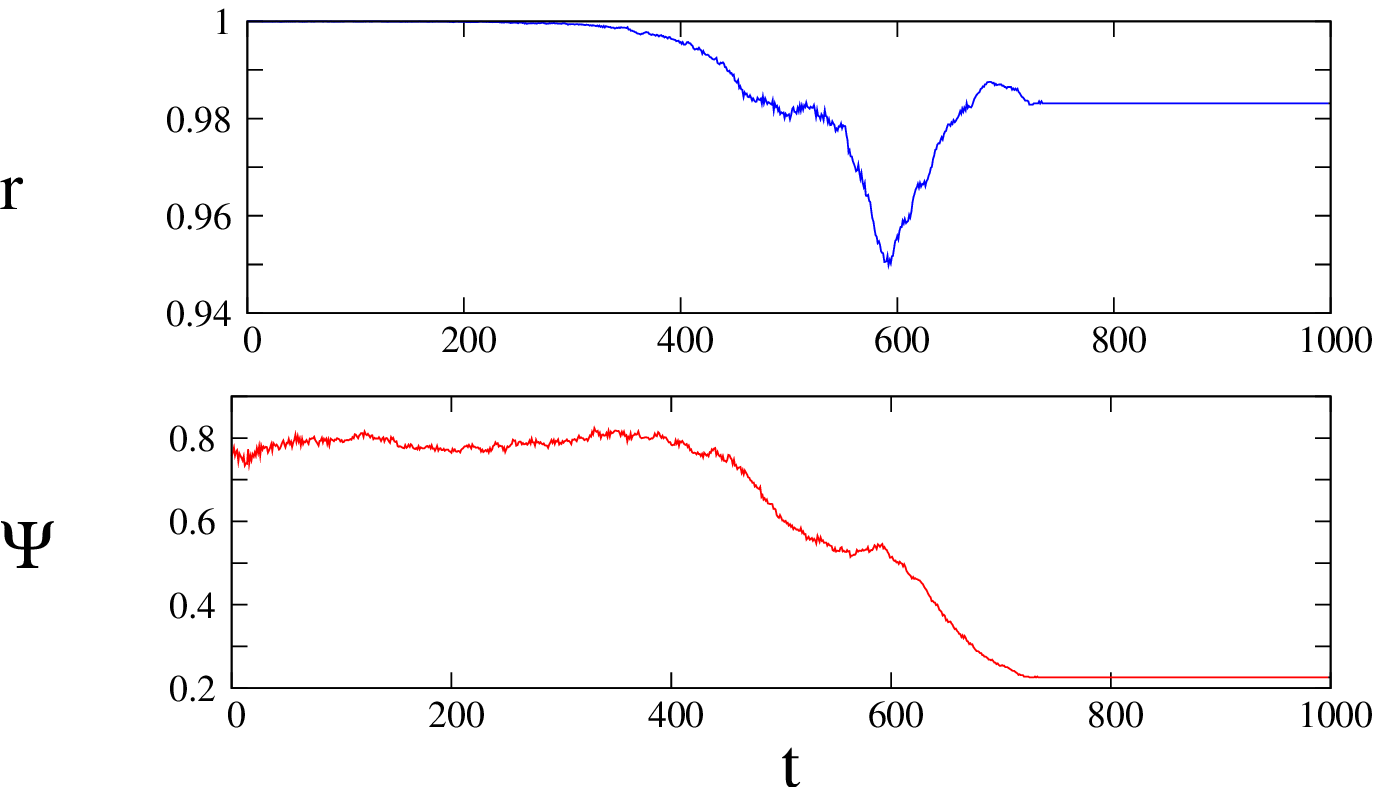}&
\hspace{1.5cm}
(b)&
\includegraphics[width=7.5cm,height=7cm]{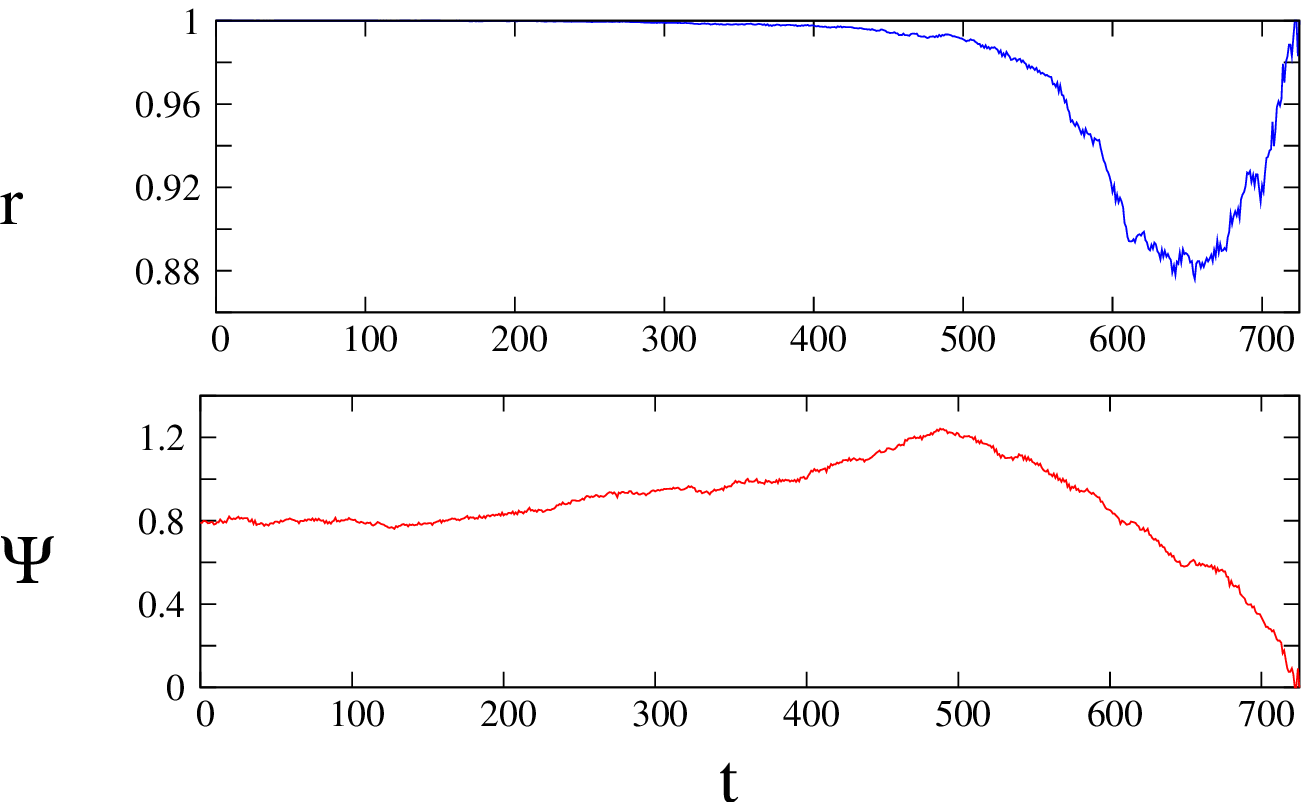}\\
\end{tabular}
\end{center}
\caption{\label{fig:globalbsl}(Color online) Plot of order parameter $r$
and average phase $\psi$ as a function of time for (a) the baseline
mechanism and (b) the $gradient$ mechanism. A total number of 2000 messages are traveling simultaneously on a $100\times 100$ lattice with 4\% hub density.\\}
\end{figure*}

We plot the order parameter $r$
and average phase $\psi$ as a function of time for the baseline
mechanism in Fig. \ref{fig:globalbsl}(a) and the gradient connections in Fig. \ref{fig:globalbsl}(b).
It is clear      
that the order parameter goes to one up to the time $t_{synch}\simeq 300 $
indicating that the queues at all the top $5$ hubs synchronize up to this point. As         
discussed earlier, this point is also the time at which the network         
congests. Thus the intimate connection between congestion and 
synchronization is clearly demonstrated by the order parameter\footnote{The value of $r$ is seen to increase at the end of the decongestion phase. This is due to the fact that for the gradient mechanisms all queues are cleared and thus take the value zero at the end of the run.  For the baseline mechanism the queues of $4_{th}$ and $5_{th}$ hubs are cleared while rest are trapped leading to a constant value of $r$ which is less than one at the end of the run.}. It is to be noted that the order parameter $r$ and average phase $\psi$ are calculated for hubs of comparable CBC values (in this case the top five hubs).

\subsection{Other decongestion schemes}

Decongestion schemes based on random assortative connections between the
top five hubs ($CBC_a$(one way) and $CBC_c$ (two way)) and the top $5$
hubs and randomly chosen other hubs ($CBC_b$(one way)  and 
$CBC_d$ (two way)) have also proved to be effective. The phenomena 
of complete synchronization and phase synchronization can be seen for
these schemes as well. (See Table~\ref{tab:table3}). Apart from the gradient mechanism complete synchronization is seen for the $CBC_{c}$ mechanism as well, where the $4_{th}$ and $5_{th}$ ranked hubs (ranked by $CBC$) are completely synchronized. Unlike the gradient mechanism, the $4_{th}$ and $5_{th}$ ranked hubs have a direct two way connection for this realization of the  $CBC_{c}$ mechanism. Both these hubs have comparable CBC values (See Table~\ref{tab:table1}) and therefore, we see that  the queue lengths are completely synchronized. The error to the fit to the $y=x$ line is 0.926. The FTLE of the queue lengths of these hubs is less than zero indicating complete synchronization. No complete synchronization is observed for the other assortative mechanisms. Its clear from Table~\ref{tab:table3} that the top most hub (labeled x) drives the rest of the top five hubs. Global synchronization emerges for these cases as well. Thus it is seen that synchronization in queue lengths is a robust phenomena. Irrespective of the nature of connections between high CBC hubs, synchronization in queue lengths of highly congested hubs exists during the congested phase.

\begin{table*}
\caption{\label{tab:table3}The table shows the pair of queue lengths of
the top five hubs which are completely synchronized (with errors to the fit to the $y=x$ line  in []-s) and phase
synchronized for a typical configuration. Similar results are obtained for different configurations as well. We consider a ${100\times 100}$ lattice with 4$\%$ hub
density and $D_{st}$ = 142 . A total number of 2000 messages are traveling simultaneously in the lattice. The run time is set at
$4D_{st}$.}
\begin{ruledtabular}
\begin{tabular}{ccc}
Mechanism & Complete Synchronization & Phase Synchronization\\
\hline
$CBC_{a}$ & - & (x,y),(x,z),(x,u),(x,v),(y,z),(y,u),(y,v),(z,u),(z,v),(u,v) \\
$CBC_{b}$ & - & (x,y),(x,z),(x,u),(x,v),(y,z),(y,u),(y,v),(z,u),(z,v),(u,v) \\
$CBC_{c}$ & (u,v)[0.926] & (x,y),(x,z),(x,u),(x,v),(y,z),(y,u),(y,v),(z,u),(z,v) \\
$CBC_{d}$ & - & (x,y),(x,z),(x,u),(x,v),(y,z),(y,u),(y,v),(z,u),(z,v),(u,v) \\
Gradient  & (y,z)[0.992] & (x,y),(x,z),(x,u),(x,v),(y,u),(y,v),(z,u),(z,v),(u,v)\\
\end{tabular}
\end{ruledtabular}
\end{table*}

\section{A network with random Waxman topology}

\begin{figure*}                          
\begin{center}
\includegraphics[width=17.5cm,height=7.5cm]{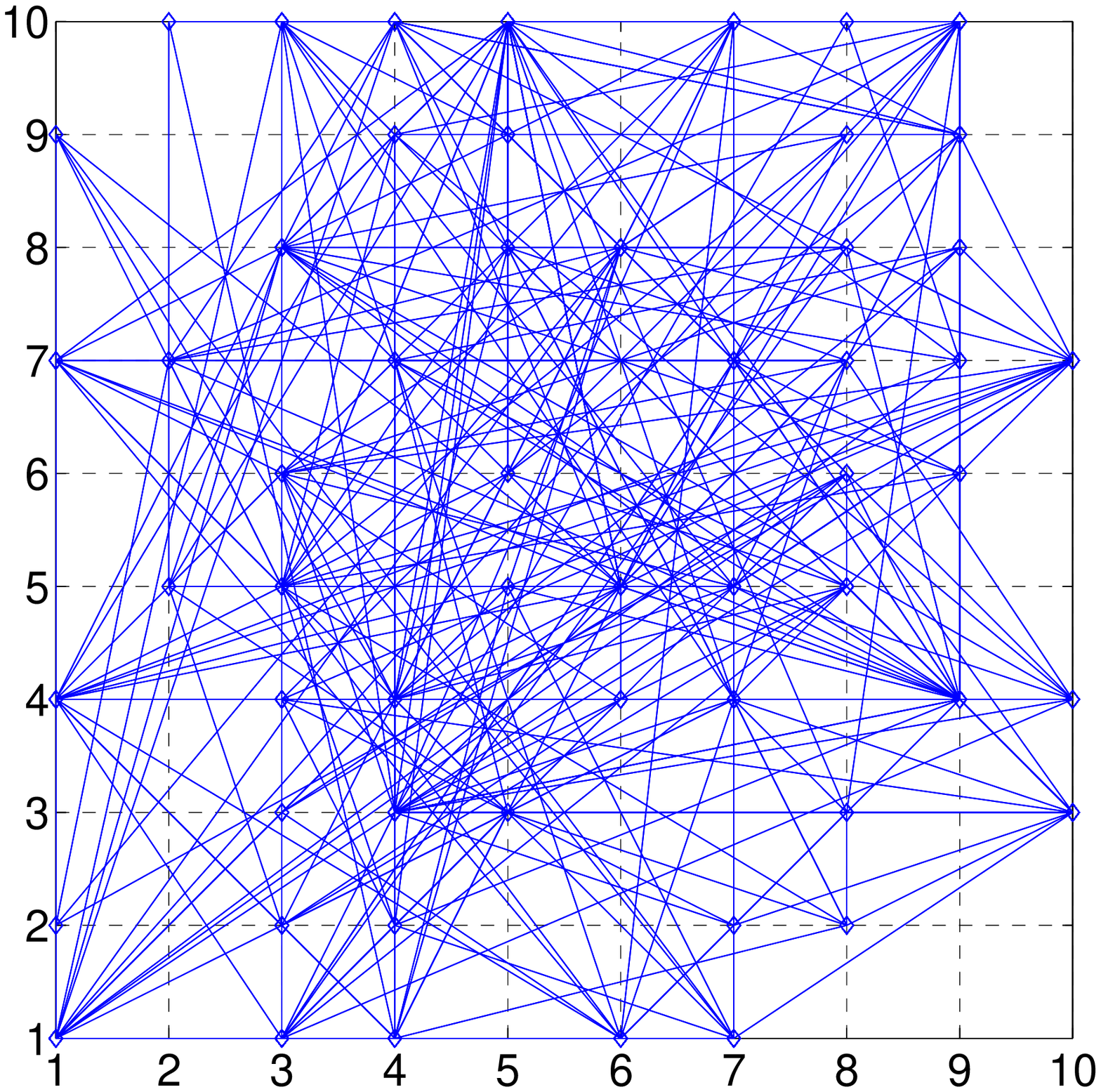}
\end{center}
\caption{\label{fig:waxfig}(Color online) Figure shows a Waxman topology network generated by connecting $55$ points by the Waxman algorithm for $\alpha=0.05$ and $\beta=0.1$, on a $10\times 10$ lattice.  The number of links increases as the values of $\alpha$ and $\beta$ are increased.\\}
\end{figure*}             

The random network topology generator introduced by Waxman \cite{waxmangraph} is a geographic model for the growth of a computer network. In this model the nodes of the network are uniformly distributed in the plane and edges are added according to probabilities that depend on the distances between the nodes.
Such networks are useful for Internet modeling due to the distance dependence in link formation which is characteristic of real world networks \cite{lakhina} and have  been widely used to model the topology of intra-domain networks \cite{verma,neve,shin,guo}.  
We study  queue length synchronization on this network, and compare it  with 
the synchronization seen for the geographically clustered network of Section II. We consider the case where the Waxman graphs are generated on a rectangular coordinate grid of side $L$ with a probability $P(a,b)$ of an edge from node $a$ to node $b$ given by

\begin{equation}
P(a,b)=\beta\exp(-\frac{d}{\alpha M})
\end{equation}
where the parameters $0< \alpha,\beta < 1$, $d$ is the Euclidean distance from $a$ to $b$ and $M=\sqrt2\times L$ is the maximum distance between any two nodes \cite{waxmangraph,naldi,waxsquare}. Larger values of $\beta$ results in graphs with larger link densities and smaller values of $\alpha$ increase the density of short links as compared to the longer ones.

Here we select a $100 \times 100$ lattice. A topology similar to Waxman graphs is generated by selecting randomly a pre-determined number $N_{w}$ of nodes in the lattice. The nodes are then connected by the Waxman algorithm, resulting in a topology which is similar to Waxman graphs (Fig. \ref{fig:waxfig}). Additionally, each node has a connection to its nearest neighbors. We study message transfer by the same routing algorithm as used in Section II. We evaluate the coefficient of betweenness centrality of the nodes and select the five top most nodes ranked by their CBC values. We compare the synchronization in queue lengths of these nodes for different values of $\beta$ and $\alpha$ for simultaneous message transfer. 

\begin{figure*}                                               
\begin{center}                                                
\begin{tabular}{cccc}    
(a)&
\includegraphics[width=7.5cm,height=7cm]{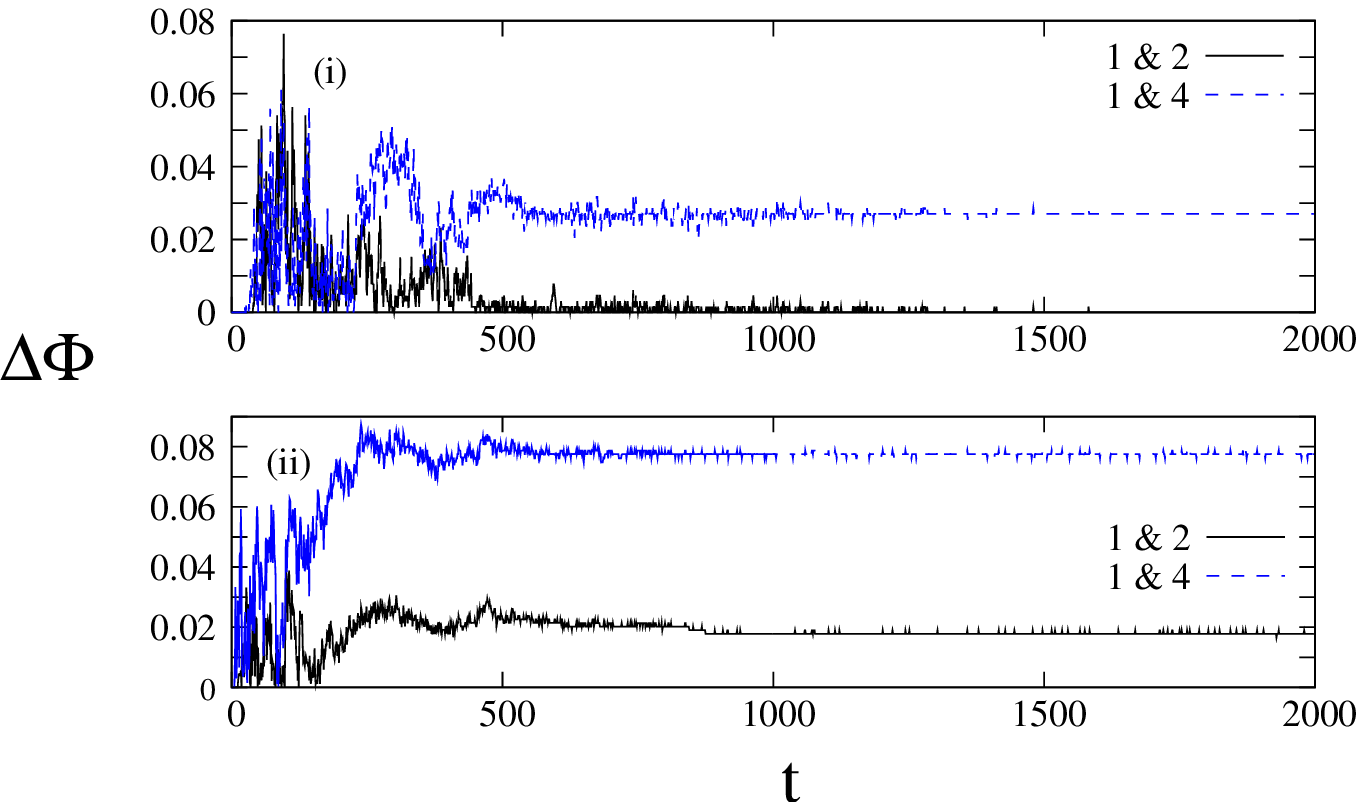}&      
\hspace{1.5cm}  
(b)&
\includegraphics[width=7.5cm,height=7cm]{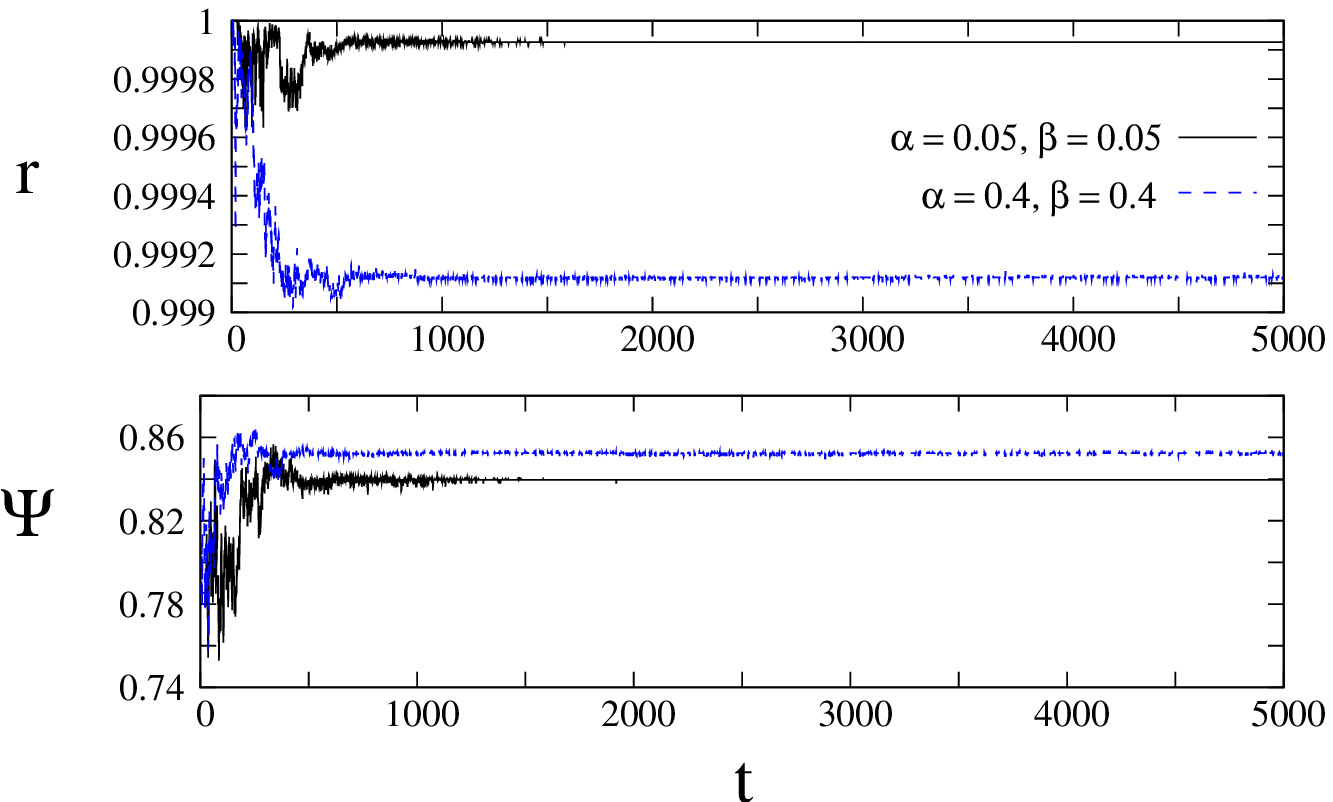}\\   
\end{tabular}                                                 
\end{center}                                                  
\caption{\label{fig:waxphasesynch}(Color online)  (a) Plot of  phase synchronization in queue length of the top most congested nodes ranked by their CBC values for (i) $\alpha=0.05$ and $\beta=0.05$; (ii) $\alpha=0.4$ and $\beta=0.4$ for simultaneous message transfer. $N_{w}=100$ points are chosen randomly in the lattice and connected by Waxman algorithm. We drop $N=2000$ messages simultaneously on the lattice. (b) The plot of global synchronization parameter $r$ and $\Psi$ as a function of time $t$ for $\alpha=0.05$ and $\beta=0.05$ ; $\alpha=0.4$ and $\beta=0.4$ for simultaneous message transfer.}
\end{figure*}

The phenomenon of phase synchronization in queue lengths is again studied for simultaneous message transfer where $N=2000$ messages flow simultaneously on the lattice with $N_{w}=100$ points chosen randomly in the lattice and connected by Waxman algorithm. The source target separation is $D_{st}=142$ as before, and is again the Manhattan distance between source and target. If $\alpha=0.05$ and $\beta=0.05$, the number of links between the randomly distributed nodes are very few as in  the geographically clustered network.  In such a situation, messages are cleared slowly and we observe strong phase synchronization (Fig. \ref{fig:waxphasesynch}(a)(i)). An increment in the values of $\alpha$ and $\beta$ increases the density of links. Messages are cleared faster due to the presence of a large number of short cuts which leads to larger fluctuations in phase, and weaker phase synchronization is seen (Fig. \ref{fig:waxphasesynch}(a)(ii)). For both the situations messages get trapped
and after some time and the phase gets locked. We see the saturation in the plot of $\Delta\Phi$. Global synchronization is also seen in this system (Fig. \ref{fig:waxphasesynch}(b)).

\section{Constant Density Traffic}

\begin{figure*}
\begin{center}
\begin{tabular}{cccc}
(a)&
\includegraphics[width=7.5cm,height=7cm]{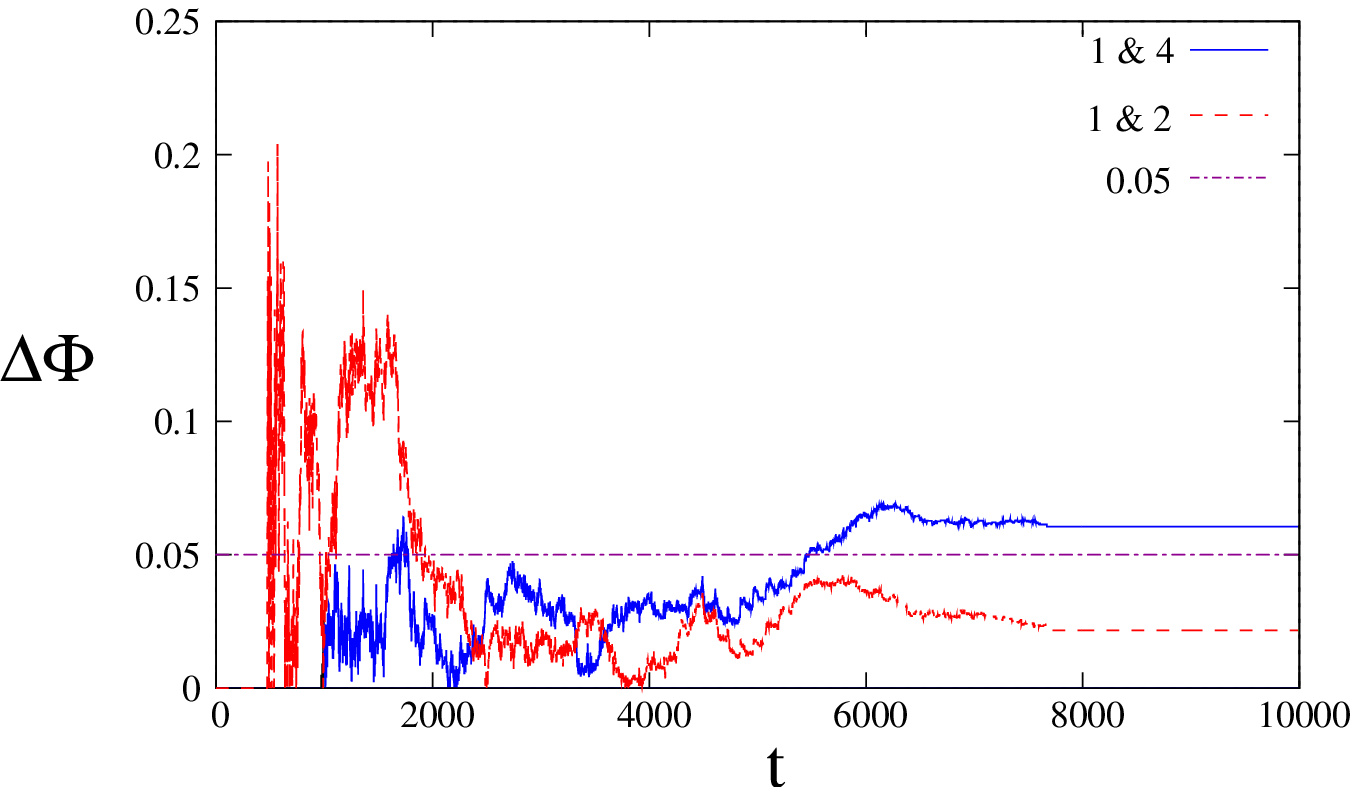}&
\hspace{1.5cm}
(b)&
\vspace{0.3cm}
\includegraphics[width=7.5cm,height=7cm]{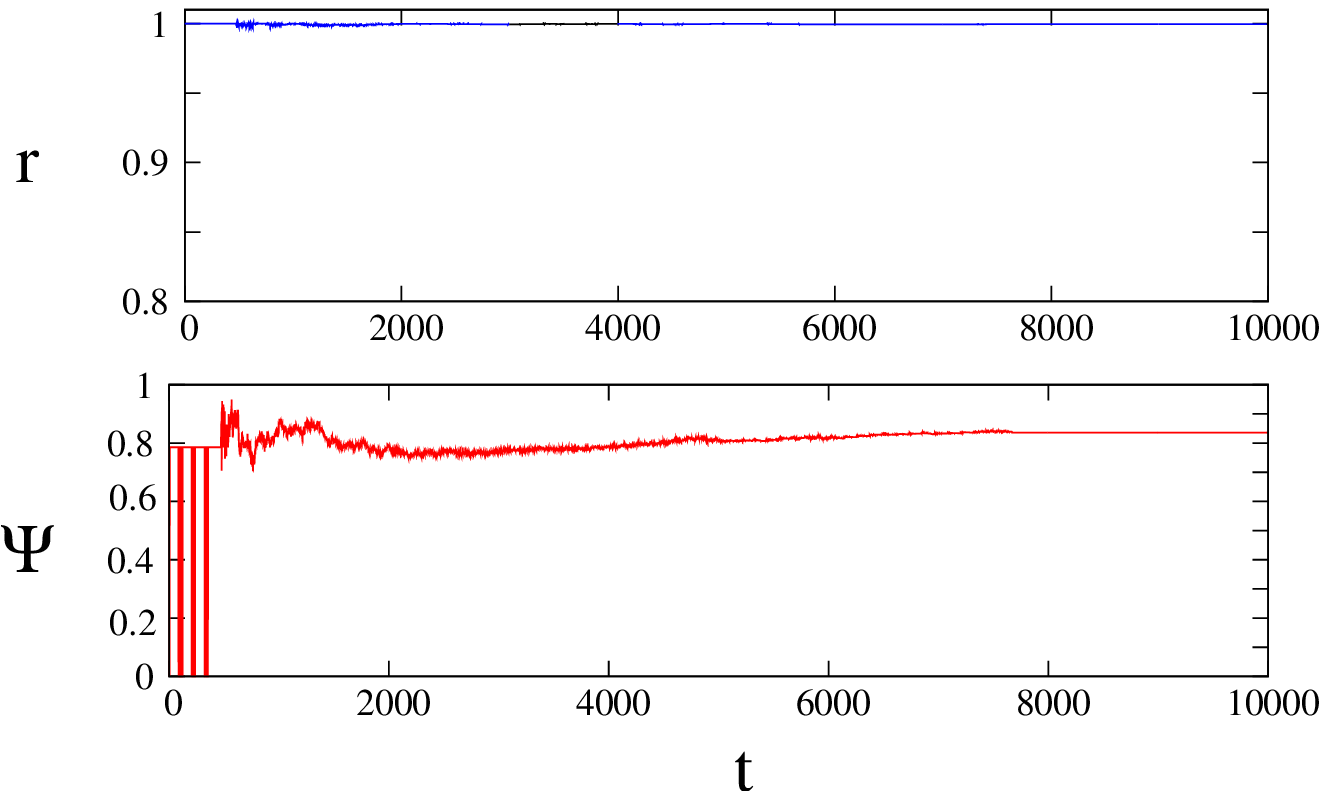}\\
(c)&
\includegraphics[width=7.5cm,height=7cm]{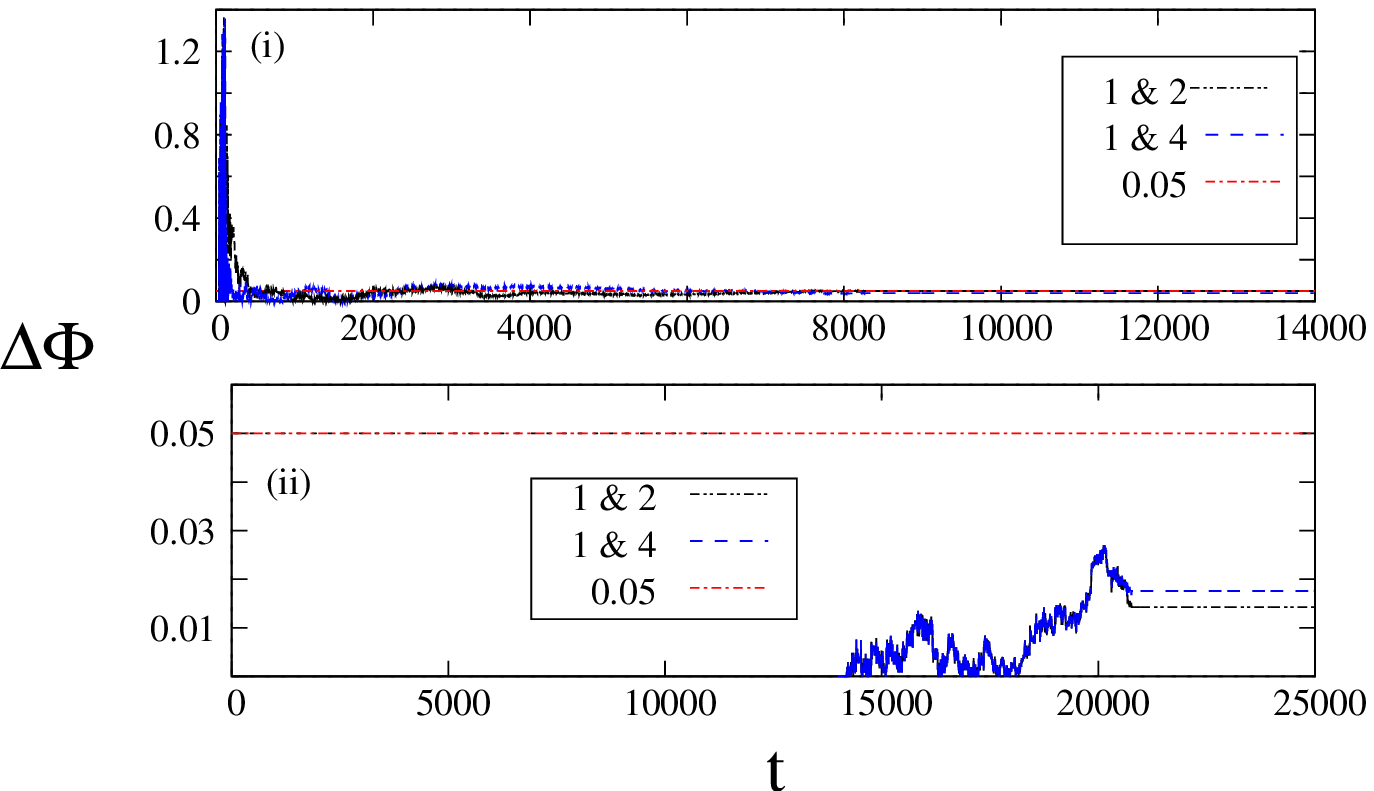}&
\hspace{1.5cm}
(d)&
\includegraphics[width=7.5cm,height=7cm]{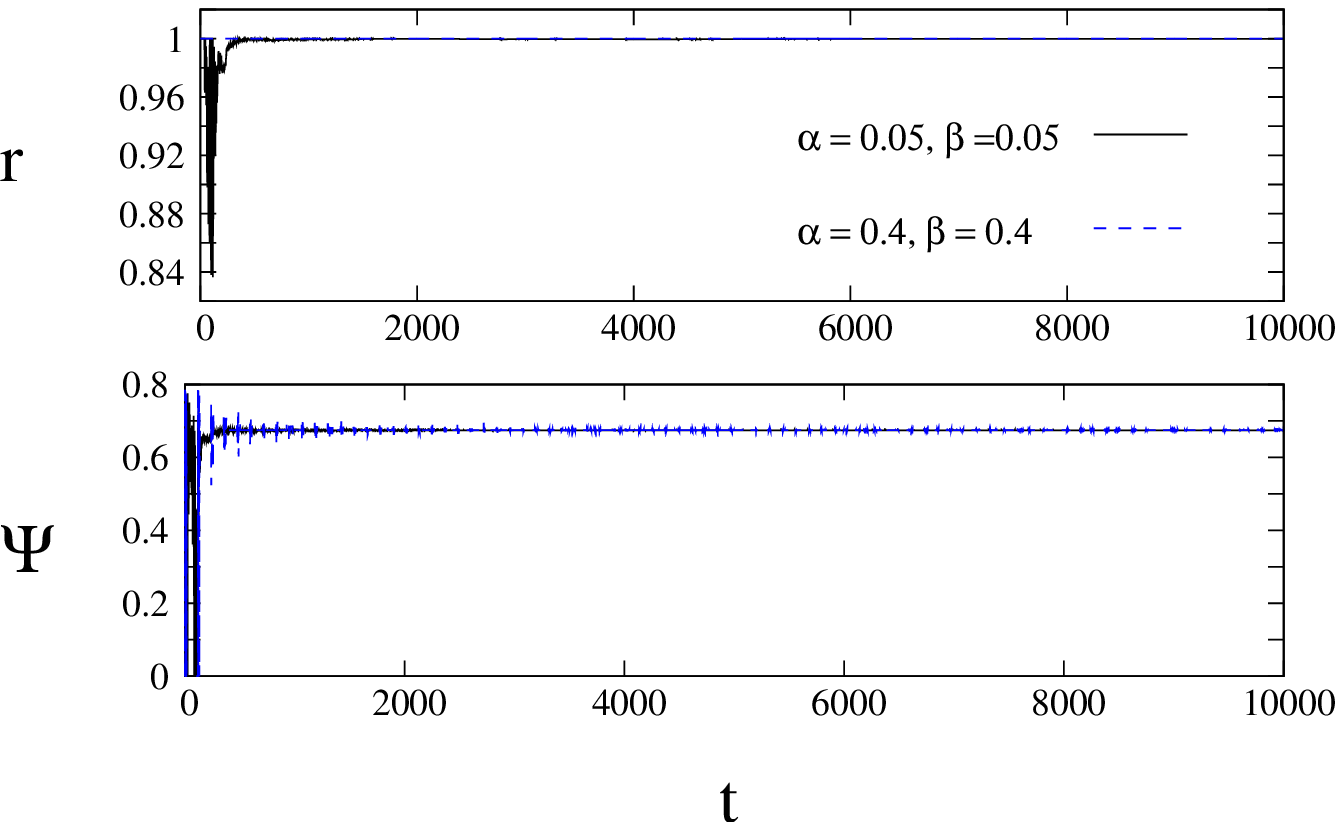}\\
\end{tabular}
\end{center}
\caption{\label{fig:cdt1}(Color online) (a) Phase synchronization for constant density traffic for pairs of hubs as labeled and (b) Plot of order parameter $r$ and average phase $\psi$ as a function of time for the baseline mechanism. Messages are fed continuously at every $120$ time steps for a total run time of $12000$. We consider 100 messages in a ${100\times 100}$ lattice with 1$\%$ hub density and $D_{st}$=142. (c) Plot of  phase synchronization in queue length of the top most congested nodes ranked by their CBC values for (i) $\alpha=0.05$ and $\beta=0.05$; (ii) $\alpha=0.4$ and $\beta=0.4$. $N_{w}=500$ points are chosen randomly in the lattice and connected by Waxman algorithm. We drop $N_{m}=100$ messages on the lattice at $120$ time steps. (d) The plot of global synchronization parameter $r$ and $\Psi$ as a function of time $t$ for $\alpha=0.05$ and $\beta=0.05$ ; $\alpha=0.4$ and $\beta=0.4$ for constant density traffic.\\}
\end{figure*} 

In the previous sections we discussed synchronization in queue lengths for simultaneous message transfer where $N$ messages are deposited simultaneously on the lattice and no further messages are fed on to the system. In this section we study synchronization in queue lengths for the constant density traffic. For the model of section II  we consider $100$ messages fed at every $120$ time steps  with 100 hubs and $D_{st}$ = 142 for a total run time of $12000$. Again, two phases, viz. the decongested phase and the congested phase are seen. In the decongested phase all messages are delivered to their respective targets, despite the fact that new messages are coming in at regular intervals. The queue lengths are not phase synchronized during this phase. In the congested phase, messages tend to get trapped in the vicinity of the hubs of high CBC, due to the reasons discussed in Section II. As more messages come in, the number of undelivered messages increase and the queue lengths start increasing
  until total trapping occurs in the system. During this phase, the creation of messages is stopped and the system attains maximal congestion. The queue lengths show phase synchronization during this phase (Fig. \ref{fig:cdt1}(a)). Initially the fluctuations in $\Delta\Phi$ are large. After a time $t\simeq 2000$ the queue lengths start increasing and the fluctuations are reduced indicating stronger phase synchronization. As soon as maximal congestion takes place ($\simeq 7800$), the phase difference attains a constant value. Global synchronization is also seen in this system as can be seen from 
Fig. \ref{fig:cdt1}(b). Note that the scales on which the phase difference and the global synchronization parameter fluctuate is very small indicating a much stronger version of synchronization than in the earlier case.

The results are compared with constant density traffic for the Waxman topology network as discussed earlier. We consider 100 messages fed continuously at every $120$ time steps for a total run time of $50000$ with $N_{w}=500$ points and $D_{st}$ = 142. If $\alpha=0.05$ and $\beta=0.05$ messages get trapped in the system very fast ($t_{c}=8000$). Phase synchronization in queue lengths is observed in such cases (Fig. \ref{fig:cdt1}(c)(i)). If the values of $\alpha$ and $\beta$ increase the number of links increase. Phase synchronization in queue lengths take place at much higher time for $(i)$ $\alpha=0.05$ and $\beta=0.4$ and $(ii)$ $\alpha=0.4$ and $\beta=0.05 $. If $\alpha=0.4$ and $\beta=0.4$ the density of links is very large and all the top five nodes have approximately equal queue lengths. Hence we observe a stronger phase synchronization where the fluctuation of $\Delta\Phi$ is well below the predetermined constant $C=0.05$ (Fig. \ref{fig:cdt1}(c)(ii)). Global synchronization is also observed in this system (Fig. \ref{fig:cdt1}(d)).

Thus, the model networks studied here show phase synchronisation as well as 
global synchronisation in the congested phase. The two traffic patterns studied here are those of a single time deposition, and that of constant density traffic. Real life networks can have traffic patterns which wax and wane several times in a single day. However, synchronisation phenomena can be seen in real networks as well. We demonstrate this phenomenon  
in terms of the number of views at different web-sites, in the next section. 

\section{Synchronization for real traffic data}

\begin{figure*}                                              
\begin{center}                                                
\includegraphics[width=8.5cm,height=7.5cm]{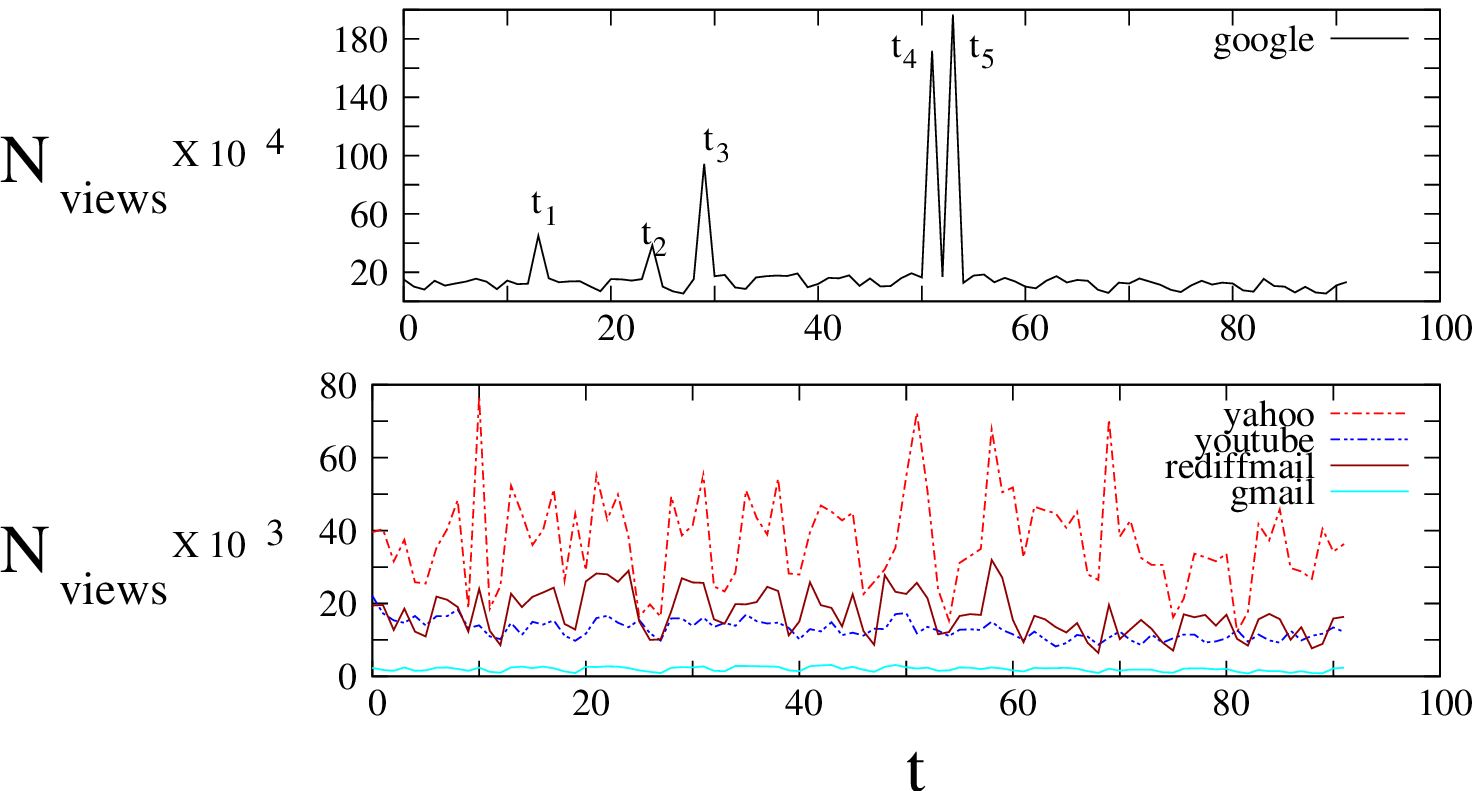}
\end{center}                                                  
\caption{\label{fig:nviews}(Color online) The plot of number views for a given website per day is obtained from $proxy2.iitm.ac.in$ for a period of $92$ days, from $01/10/2008$ to $31/12/2008$ (denoted by $t$ in x-axis). The websites are $www.google.com$, $www.gmail.com$, $www.yahoo.com$, $www.youtube.com$ and $www.rediffmail.com$. The data is counted specifically for the given sites and not for sub domains.}
\end{figure*}

We discuss phase synchronization in the Internet traffic data in the Indian Institute of Technology Madras, India. The data is collected for the number of views to different websites. The websites are $www.google.com$, $www.gmail.com$, $www.yahoo.com$, $www.youtube.com$ and $www.rediffmail.com$. The data is counted specifically for the given sites and not for subdomains \footnote{The data is obtained from the log files (generated by the SQUID software) of the proxy server in the IIT Madras campus}. Fig. \ref{fig:nviews}(a) shows the total number of views for the five websites per day for a period of $92$ days ,from $01/10/2008$ to $31/12/2008$ ($t$ in days on the x-axis). In Fig. \ref{fig:nviews}(b) we plot the number of views per minute for $10^{th}$ November  $2008$ ($t$ in minutes on the x-axis).

\begin{figure*}                                               
\begin{center}                                                
\begin{tabular}{cccc}    
(a)&                                       
\includegraphics[width=7.5cm,height=7cm]{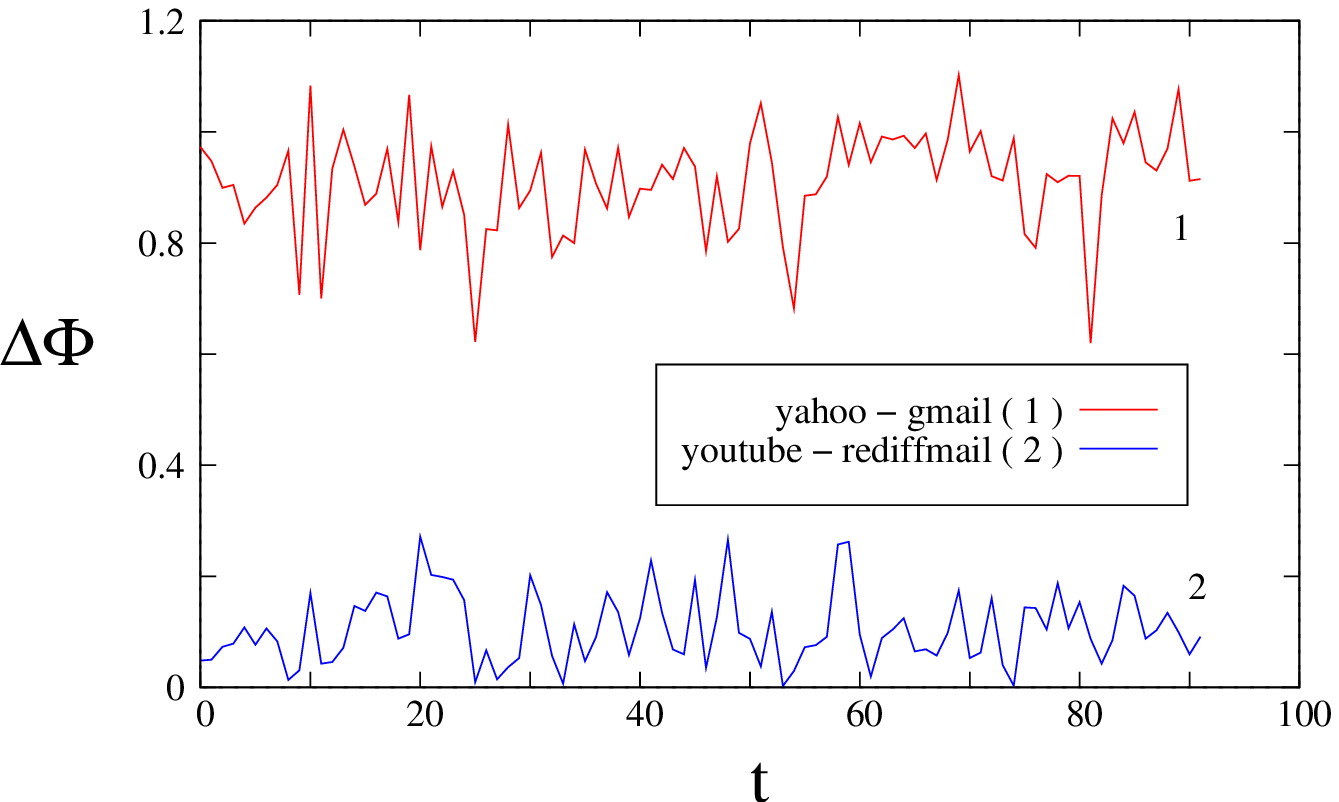}&
\hspace{1.5cm}
(b)&
\vspace{0.3cm}
\includegraphics[width=7.5cm,height=7cm]{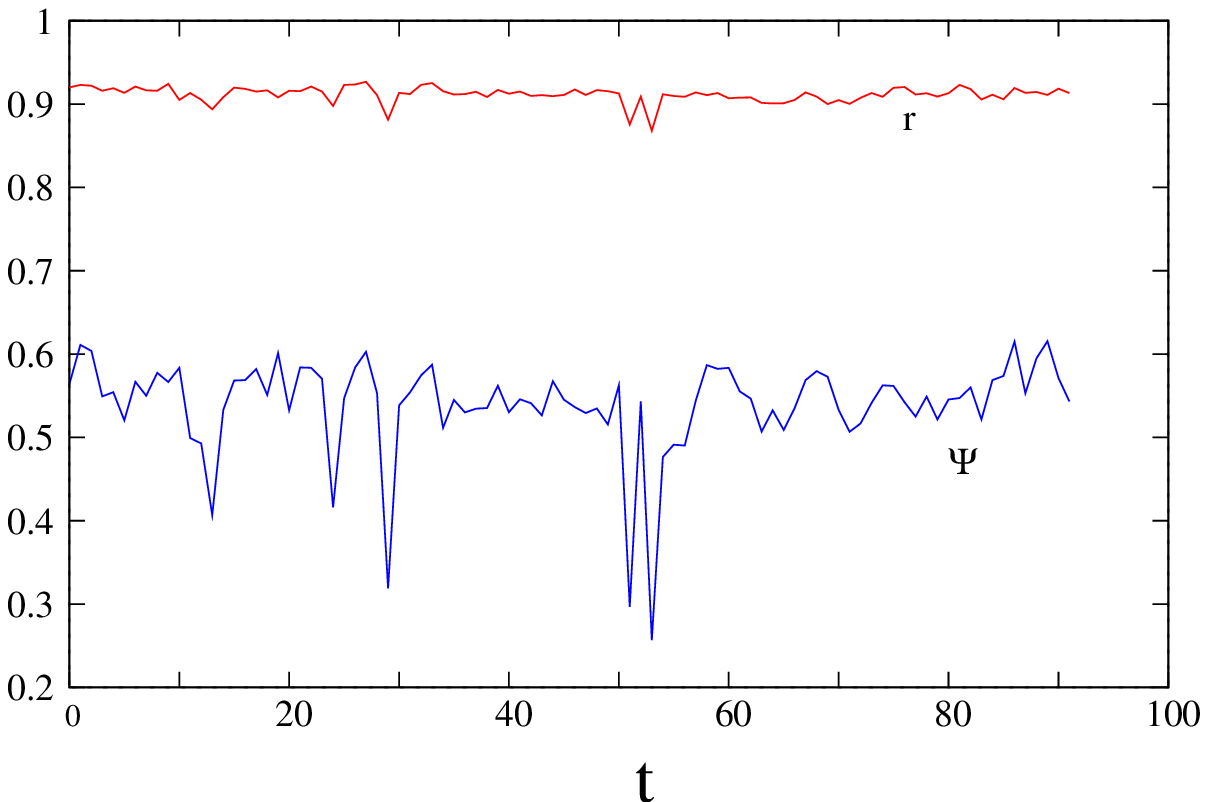}\\
(c)&
\includegraphics[width=7.5cm,height=7cm]{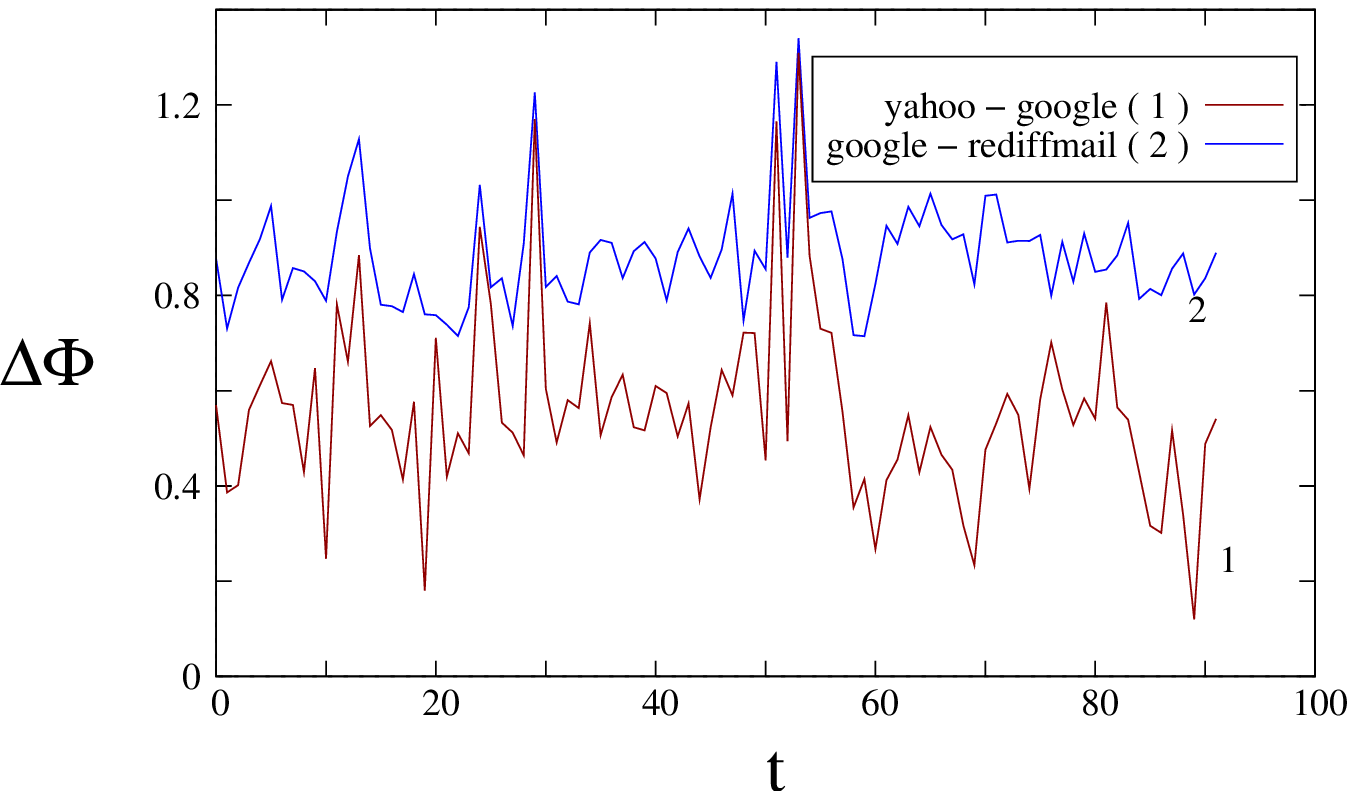}&
\hspace{1.5cm}      
(d)&
\includegraphics[width=7.5cm,height=7cm]{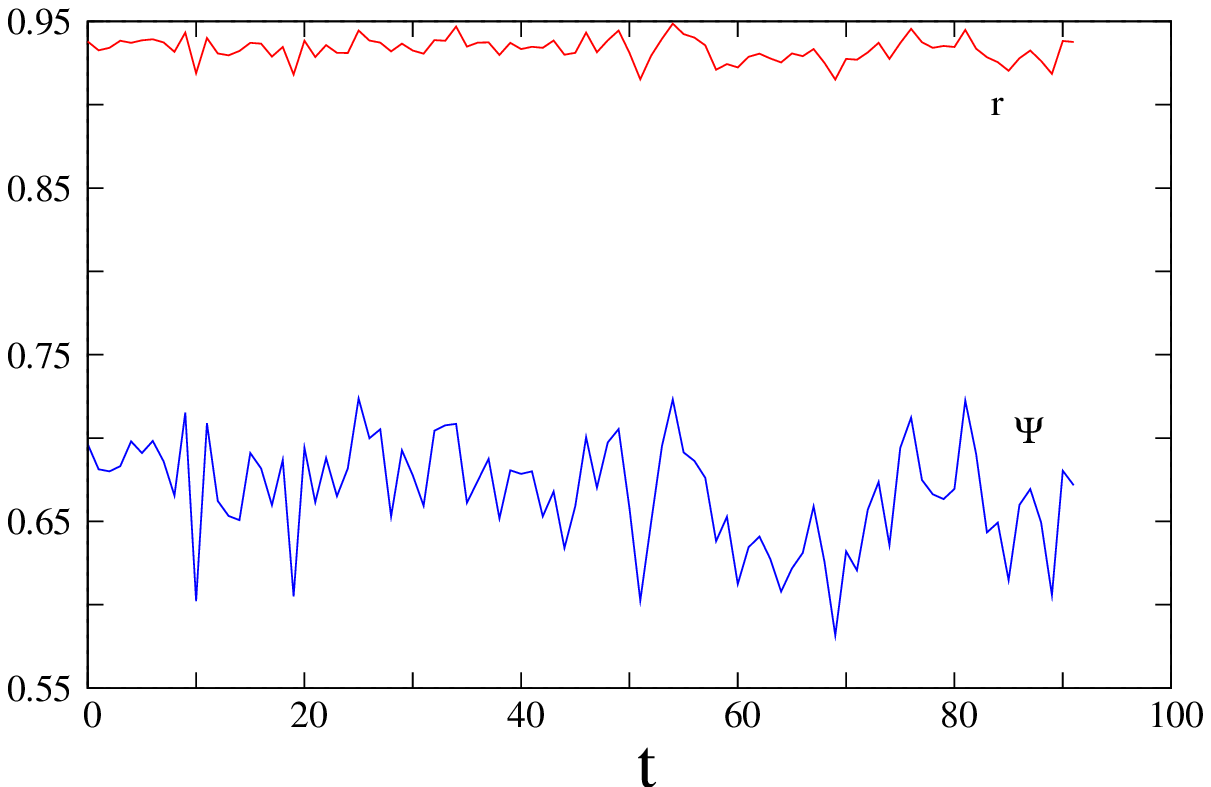}\\   
\end{tabular}                                                 
\end{center}                                                  
\caption{\label{fig:synchreal}(Color online)  (a) Phase synchronization in number of views for the pairs of websites (yahoo, gmail) and (youtube, rediffmail). (c) No phase synchronization is observed for (yahoo, google) and (google, rediffmail) as indicated by the large fluctuations in $\Delta\Phi$. (b) Large fluctuations in global synchronization parameter $r$ and $\Psi$ are seen when all the five websites are taken into account. (d) The fluctuations are reduced when the website $www.google.com$ is not taken into account.}
\end{figure*}

As is evident from the plots the number of views for the website google is very large as compared to the rest. In Fig. \ref{fig:nviews} it is seen that $N_{views}$ for google show abrupt high peaks at $t_{1}=13$ ($13^{th}$ October), $t_{2}=24$ ($24^{th}$ October), $t_{3}=28$ ($28^{th}$ October), $t_{4}=51$ ($20^{th}$ November) and $t_{5}=53$ ($22^{nd}$ November) (See \footnote{ $13^{th}$ October, $24^{th}$ October, $20^{th}$ November and $22^{nd}$ November were all dates for semester examinations in  the Indian Institute of Technology Madras, India and students tend to access google more for tutorials and solutions available in the web, at these times. Again $28^{th}$ October  is the festival of $Diwali$ which is a national holiday in India. The Internet users on campus appear to have  spent most of this holiday  browsing. The value of  $N_{views}$ for all the websites reach their peak during the day time but decreases during night, contrary to the notion that web browsing increases during the night. This is due to the fact that Internet is unavailable 
between 20:00 hours to 04:00 hours in the student hostels.}).

\begin{figure*}                                             
\begin{center} 
\begin{tabular}{cccc}  
(a)&
\includegraphics[width=7.5cm,height=7cm]{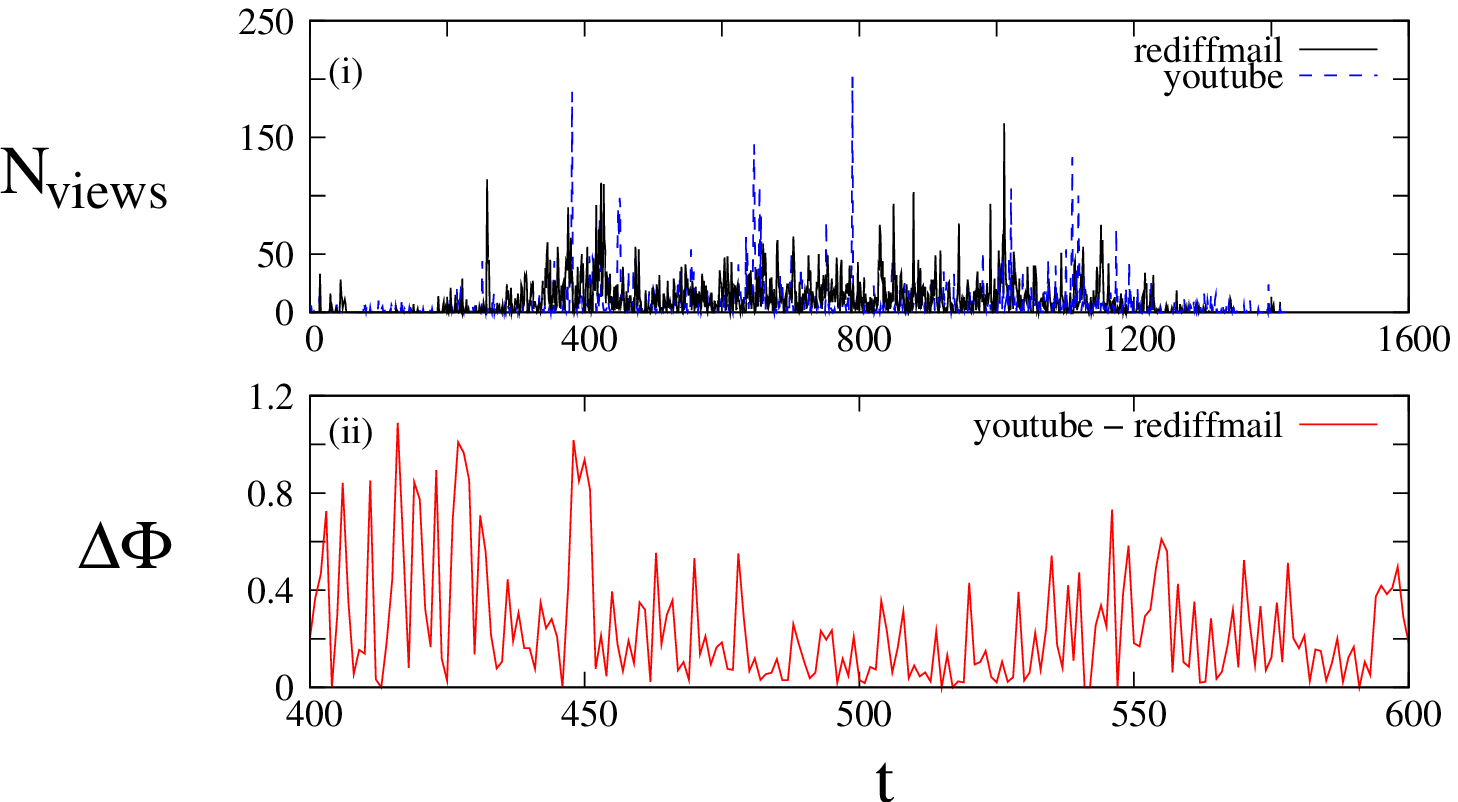}&
\hspace{1.5cm}
(b)&
\includegraphics[width=7.5cm,height=7cm]{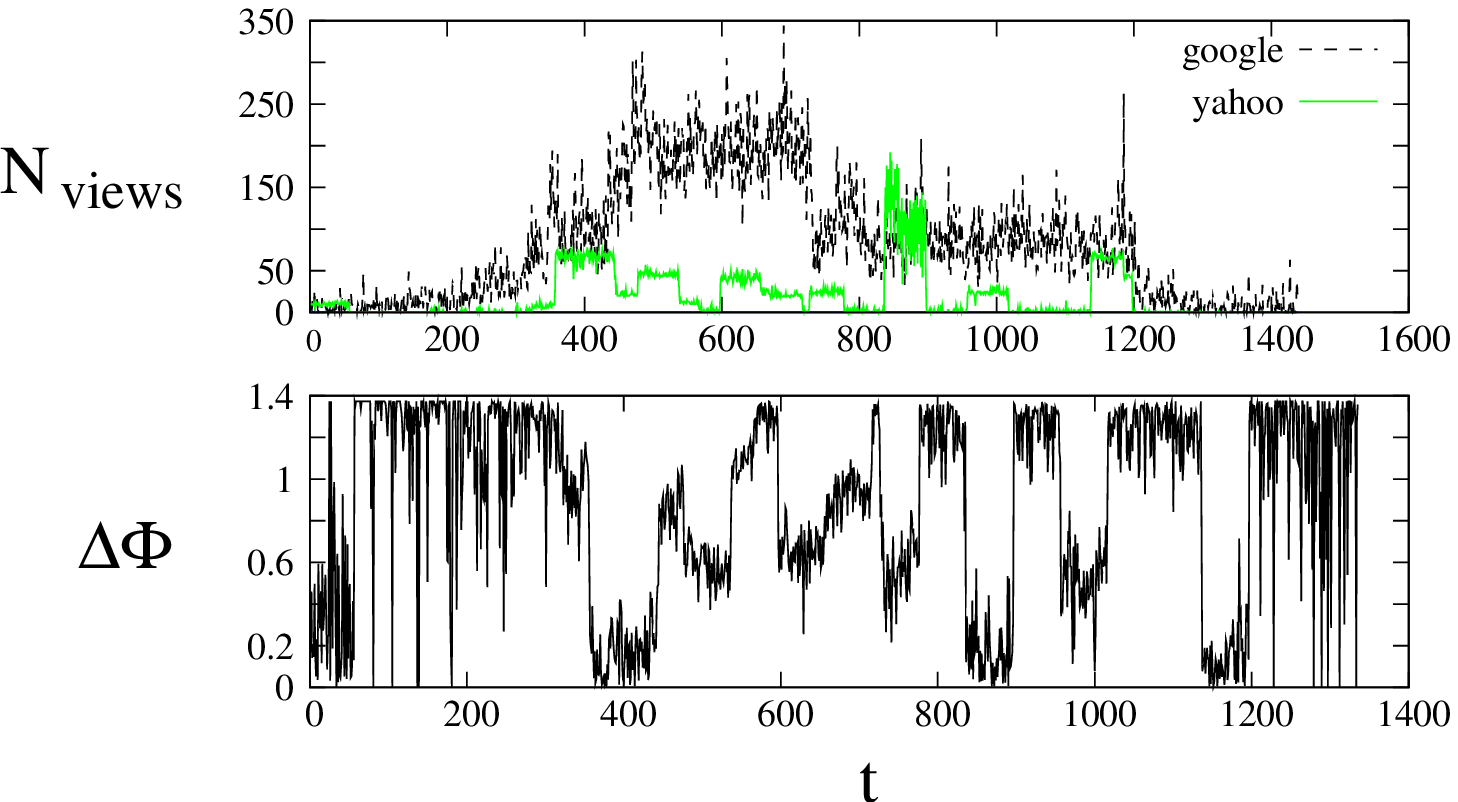}\\ 
\end{tabular}                                                 
\end{center}                                                  
\caption{\label{fig:phasesynch1}(Color online) (a(i)) The plot of number of views per minute (denoted by $t$ in x-axis) for $10^{th}$ Nov $2008$. The websites are $www.youtube.com$ and $www.rediffmail.com$. (a(ii)) Phase synchronization in number of views is observed for the pair (youtube, rediffmail) in the time interval $t=400-600$. (b) The websites shown are $www.google.com$ and $www.yahoo.com$. We observe intermittent phase synchronization in the number of views for (yahoo, google). The data is counted specifically for the given sites and not for sub domains.}
\end{figure*}

In Fig. \ref{fig:synchreal} we plot the phase synchronization in number of views for different websites. The phase is defined as in Eq.1. It is observed that the phase locking condition holds for the pairs (yahoo, gmail) and (youtube, rediffmail) (See Fig. \ref{fig:synchreal}(a)). No such phase locking condition exists between the pairs (google, yahoo) and (google, rediffmail) (See Fig. \ref{fig:synchreal}(c)). This is due to two facts. First the number of views for $www.google.com$ are much higher than those of the other websites. Secondly, the presence of abrupt peaks for google leads to larger fluctuations. The plot of global synchronization parameters $r$ and $\Psi$ shows that the websites are synchronized in terms of number of views (See Fig. \ref{fig:synchreal}(b, d)). Larger fluctuations in $r$ and $\Psi$ are seen when all the five websites are taken into account (See Fig. \ref{fig:synchreal}(b)). The fluctuations are reduced when the website $www.google.com$ is not taken into account (See Fig. \ref{fig:synchreal}(d)). 

\begin{figure*}                                             
\begin{center} 
\begin{tabular}{cccc}  
(a)&
\includegraphics[width=7.5cm,height=7cm]{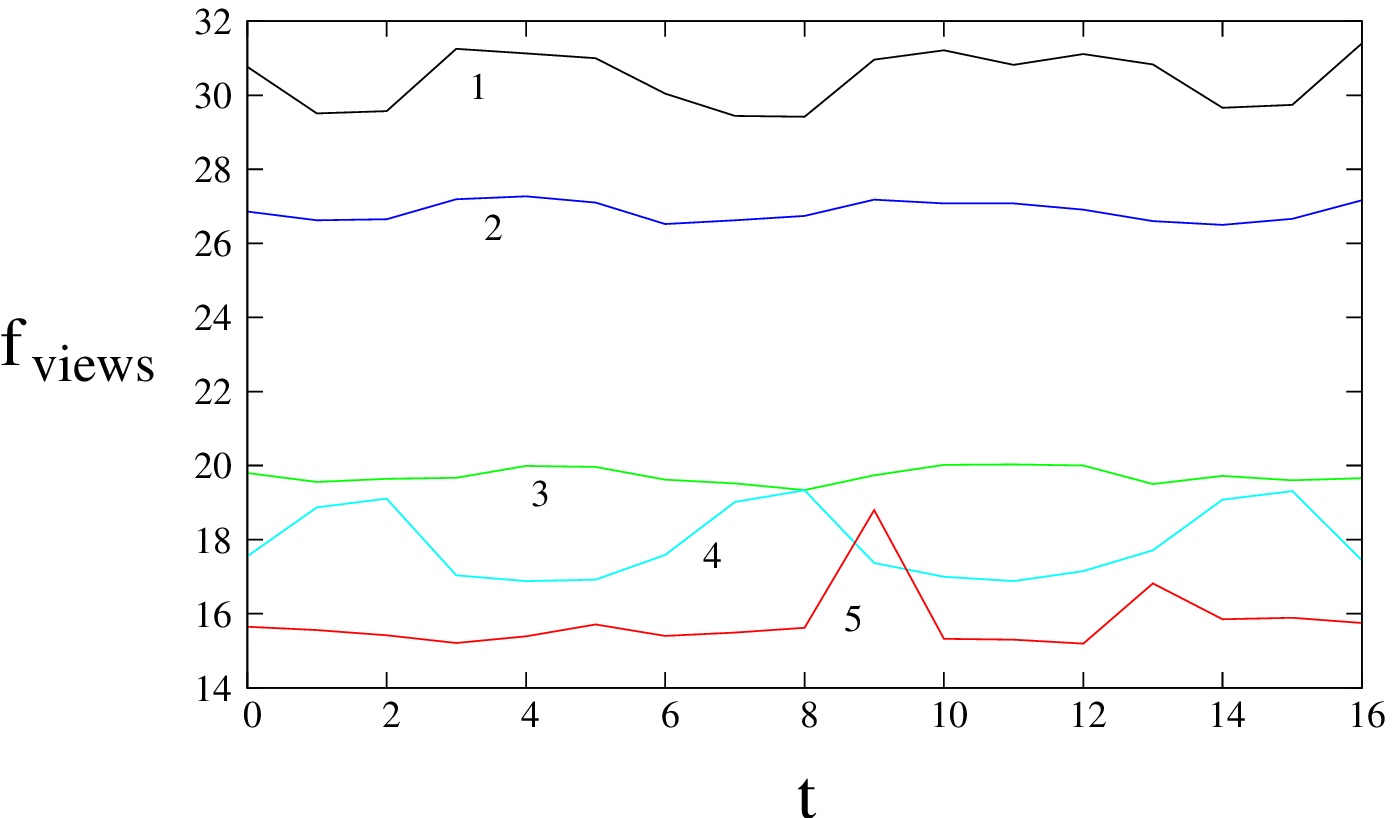}&
\hspace{1.5cm}
(b)&
\includegraphics[width=7.5cm,height=7cm]{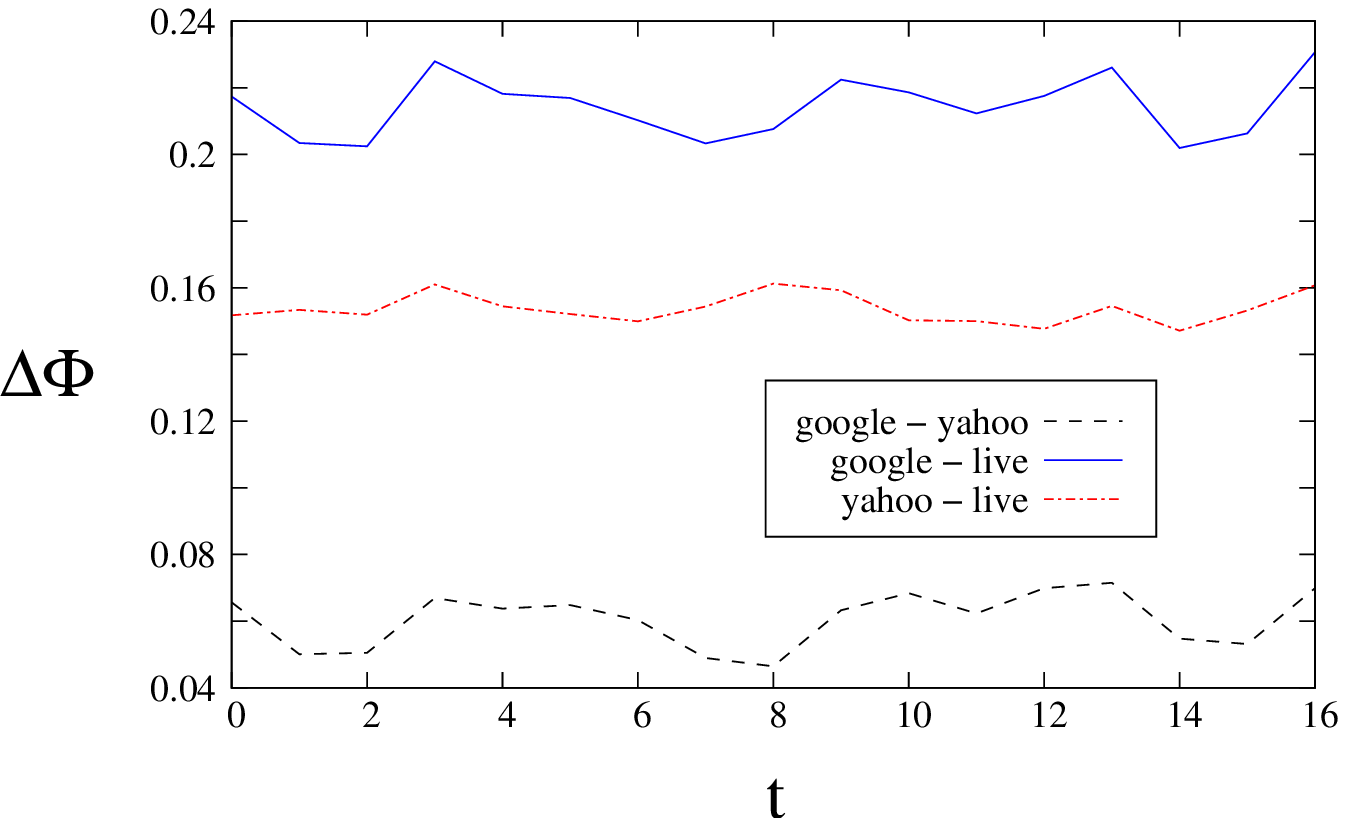}\\
\end{tabular}                                                 
\end{center}                                                  
\caption{\label{fig:phasesynch2}(Color online) (a) The percentage of global Internet users ($f_{views}$) who visit the respective websites listed for a period of $16$ days from 08/02/2009 to 24/02/2009 (denoted by $t$ in x-axis). The websites are labeled as 1 ($www.google.com$), 2 ($www.yahoo.com$), 3 ($www.youtube.com$), 4 ($www.live.com$) and 5 ($www.msn.com$), where the numbers denote the global rank of the websites. (b) Phase synchronization in $f_{views}$ is observed for pairs of websites.}
\end{figure*}

 We observe similar behavior when the data is studied for $10^{th}$ Nov $2008$. It is observed that the two websites (youtube, rediffmail) peak together during the time interval $t=400 - 600$ (Fig. \ref{fig:phasesynch1}(a)). During this time interval the two websites are synchronized as shown in Fig. \ref{fig:phasesynch1}(a(ii)). As soon as the values of $N_{views}$ start decreasing for both the sites, phase synchronization is lost. When we compared the sites google and yahoo it was observed that for yahoo, the number of views increases intermittently. Hence intermittent behavior of phase synchronization is observed for yahoo and google (See Fig. \ref{fig:phasesynch1}(b)). No phase synchronization is observed between google and rediffmail. This is similar to the absence of phase synchronization in queue lengths between higher CBC hubs and hubs of low CBC values as in Fig. \ref{fig:randhub} for the $2-d$ communication network. Also we observed that hubs of comparable CBC phase synchronize. Similarly in the Internet data, websites of comparable volume of traffic phase synchronize. 

We also study the traffic data of the global top five websites ranked by the percentage of global Internet users who visit the respective website. Fig. \ref{fig:phasesynch2}(a) shows the plot of percentage of global Internet users,$f_{views}$ for the top ranked websites for a period of $16$ days from 08/02/2009 to 24/02/2009. The data has been collected from the website $www.alexa.com$. The websites are $www.google.com$, $www.yahoo.com$, $www.youtube.com$, $www.live.com$ and $www.msn.com$. Here also phase locking is observed for the pairs of websites (Fig. \ref{fig:phasesynch2}(b)).

\section{Conclusions}

To summarize, we have established a connection between synchronization and congestion 
in the case of two communication networks based on  $2-d$
geometries, a locally clustered network, and a network based on random graphs, viz. the Waxman topology network. 

We first considered the case where many messages  are deposited simultaneously on the lattice. We observed that the queue lengths of the top five hubs get phase synchronized when the system is in the congested phase i.e. the queue lengths at the hubs start piling up. Phase synchronization is lost when messages start getting delivered to their destinations and queue lengths start decreasing. Complete synchronization in queue lengths between certain pairs of top five hubs is observed when decongestion strategies are implemented by connecting the top five hubs by gradient connections or by assortative linkages. 
Phase synchronization is also seen between nodes of comparable CBC in the Waxman
topology despite the fact that the degree distribution of the Waxman graphs is entirely different from that of the locally clustered network\footnote{the geographically clustered network has a bimodal degree distribution\cite{BrajNeel},whereas the Waxman graph has degree distributions which depend on the values of $\alpha$ and $\beta$\cite{skim}. }.
The phenomenon  of synchronization in queue length in the congested phase is thus a robust one. We also observed that the phase synchronization does not exist between hubs of widely separated CBC-s, as can be seen from the lack of synchronization between the high   CBC hubs and any randomly chosen hub (which turn out to have lower CBC values). 

It is has been seen, for typical configurations,  that the central most region of the lattice is the
 most congested due to flow of messages from both sides of the lattice\cite{sat}. The top hubs ranked by their CBC values were also seen to be located at the central region of the lattice and were close to each other. In such cases,  the hub with the highest CBC value gets congested first and the remaining hubs follow, resulting in increase in queue lengths at the remaining hubs after a time lag  resulting in a phase difference between the queues at different hubs. This phase difference tends to a constant in the congested phase. Thus, phase synchronization in queue lengths exists pairwise for hubs of comparable CBC values. Global synchronization is also seen between hubs of comparable CBC-s. This is demonstrated by the behavior of the global synchronization parameters for the top five hubs.

Similar phase synchronized behavior is observed when messages are fed on the lattice at a constant rate. The parameters of synchronization, such as the phase difference and the global synchronization parameters, show small fluctuations in this case, indicating stronger synchronization as compared to simultaneous deposition.

Thus, synchronization is associated with the inefficient phase of the system. Similar phenomena can be found in the context of neurophysiological systems \cite{zemanova} and in computer networks \cite{flyod}. We observed that the most congested hub drives the rest. As soon as this hub decongests, the synchronization is lost. In the case of our communication network, this is usually the hub of highest CBC. Due to the master-slave relation between the most congested hub and the rest, there is cascading effect by which successive pairs lose synchronization. Cascading effects have been seen in other systems such as power grids \cite{lynch} and the Internet \cite{vespi}. It will be interesting to see if synchronization effects can be observed in these contexts.

Finally we studied the Internet traffic data obtained in Indian Institute of Technology, Madras, India as well as the global Internet data obtained from the Alexa website. Despite  the irregular pattern of traffic in this data, phase synchronization in the number of views is observed between two websites of comparable volume of traffic. Phase synchronization breaks down if the volume of traffic changes abruptly. Global synchronization is also observed for the Internet data. This data also shows the existence of intermittent synchronization.

These observations can be of great utility in the practical situations. For example, e.g. synchronization can be considered as a predictor of congestion. Synchronization can also be used to detect changes in the pattern of traffic or to detect abnormal traffic from a given hub. The hub from where the attack originates can be easily identified via a synchronization effect. Synchronization in transport may also provide information about the way in which the network is connected. Thus our study may prove to be useful in a number of application contexts.

\begin{acknowledgments}
We wish to acknowledge the support of DST, India under the project SP/S2/HEP/10/2003. We thank A. Prabhakar and S. R. Singh for sharing the Internet data of the campus.
\end{acknowledgments}

\end{document}